\documentclass[aps,prd,preprint,superscriptaddress,tightenlines,nofootinbib]{revtex4}

\usepackage{amsfonts,amsmath,amsthm,amssymb}
\usepackage{mathrsfs}
\usepackage{bm}
\usepackage{color}
\usepackage{epsfig}

\newcommand{\PRE}[1]{{#1}}   

\newcommand{\postscript}[2]{\setlength{\epsfxsize}{#2\hsize}
   \centerline{\epsfbox{#1}}}
\newcommand{\comment}[1]{}

\newcommand{\el}[1]{\label{#1}}
\newcommand{\er}[1]{\eqref{#1}}

\newcommand{\ci}[1]{}

\newcommand{\lb}{\left(}
\newcommand{\rb}{\right)}
\newcommand{\rc}{\right.}
\newcommand{\lsb}{\left[}
\newcommand{\rsb}{\right]}

\newcommand{\nn}{\nonumber \\}

\newcommand{\ba}{\begin{eqnarray}}
\newcommand{\ea}{\end{eqnarray}}
\newcommand{\be}{\begin{equation}}
\newcommand{\ee}{\end{equation}}
\newcommand{\bay}[1]{\left(\begin{array}{#1}}
\newcommand{\eay}{\end{array}\right)}

\newcommand{\zt}[1]{\rm{#1}}

\def\xa{{\alpha}}

\def\xb{{\beta}}

\def\xd{{\delta}}
\def\xD{{\Delta}}
\def\xe{{\epsilon}}

\def\xg{{\gamma}}
\def\xG{{\Gamma}}

\def\xs{{\sigma}}

\def\xt{{\theta}}

\def\CC{{\cal C}}

\def\CM{{\cal M}}

\def\CO{{\cal O}}

\usepackage[usenames,dvipsnames]{xcolor}
\definecolor{rossoCP3}{cmyk}{0,.88,.77,.40}
\begin{document}

\preprint{
\hfil
\begin{minipage}[t]{3in}
\begin{flushright}
\vspace*{-.2in}
MPP--2021--60\\
LMU-ASC 09/21\\
\end{flushright}
\end{minipage}
}

\title{ \color{rossoCP3}   Muon $\bm{g-2}$ discrepancy within D-brane string compactifications
\PRE{\vspace*{0.1in}} }

\author{{\bf Luis A. Anchordoqui}}

\affiliation{Department of Physics and Astronomy,\\  Lehman College, City University of
  New York, NY 10468, USA
\PRE{\vspace*{.01in}}
}

\affiliation{Department of Physics,\\
 Graduate Center, City University
  of New York,  NY 10016, USA
\PRE{\vspace*{.01in}}
}

\affiliation{Department of Astrophysics,\\
 American Museum of Natural History, NY
 10024, USA
\PRE{\vspace*{.01in}}
}

\author{{\bf Ignatios Antoniadis}}

\affiliation{Laboratoire de Physique Th\'eorique et Hautes \'Energies - LPTHE\\
Sorbonne Universit\'e, CNRS, 4 Place Jussieu, 75005 Paris, France
\PRE{\vspace*{.01in}}}

\affiliation{Institute for Theoretical Physics,\\ KU Leuven, Celestijnenlaan 200D, B-3001 Leuven, Belgium
\PRE{\vspace*{.01in}}}

\author{{\bf Xing~Huang}}
\affiliation{Institute of Modern Physics,\\ Northwest University, Xi'an
  710069, China
  \PRE{\vspace*{.01in}}
}
\affiliation{Shaanxi Key Laboratory for Theoretical Physics Frontiers,\\
  Xi'an 710069, China
  \PRE{\vspace*{.01in}}
}


\author{{\bf Dieter L\"ust}}

\affiliation{Max--Planck--Institut f\"ur Physik, \\ 
 Werner--Heisenberg--Institut,
80805 M\"unchen, Germany
\PRE{\vspace*{.01in}}
}

\affiliation{Arnold Sommerfeld Center for Theoretical Physics 
Ludwig-Maximilians-Universit\"at M\"unchen,
80333 M\"unchen, Germany
\PRE{\vspace{.01in}}
}

\author{{\bf Tomasz R. Taylor}}

\affiliation{Department of Physics,\\
  Northeastern University, Boston, MA 02115, USA 
 \PRE{\vspace*{.01in}}
}


\begin{abstract}\vskip 1.5mm
  \noindent Very recently, the Muon $g-2$ experiment at Fermilab has
  confirmed the E821 Brookhaven result, which hinted at a deviation of
  the muon anomalous magnetic moment from the Standard Model (SM)
  expectation. The combined results from Brookhaven and Fermilab show
  a difference with the SM prediction $\delta a_\mu = (251 \pm 59) \times 10^{-11}$ at a significance of
  $4.2\sigma$, strongly indicating the presence of new
  physics. Motivated by this new result we reexamine the contributions
  to $\delta a_\mu$ from both: {\it (i)}~the
  ubiquitous $U(1)$ gauge bosons of D-brane string theory
  constructions and {\it (ii)}~the Regge excitations of the string. We
  show that, for a string scale ${\cal O} ({\rm PeV})$, the
  contribution from anomalous $U(1)$ gauge bosons which couple to hadrons could help to reduce (though not fully
  eliminate) the discrepancy reported by the Muon $g-2$ Collaboration.
  Consistency with null results from LHC searches of new heavy vector
  bosons imparts the dominant constraint. We demonstrate that the
  contribution from Regge excitations is strongly suppressed as it was
  previously conjectured. We also comment on contributions from
  Kaluza-Klein (KK) modes, which could help resolve the $\delta a_\mu$
  discrepancy. In particular, we argue that for  4-stack intersecting D-brane models, the KK excitations of the $U(1)$ boson living on
  the lepton brane would not couple to hadrons and
  therefore can evade the LHC bounds while fully bridging the $\delta
  a_\mu$ gap observed at Brookhaven and Fermilab.
\end{abstract}

\maketitle

\section{Introduction}

The gyromagnetic factor $g$ is defined by the relation between the
particle's spin $\vec{s}$ and its magnetic moment
$\vec{\mu}=g \, e\, \vec{s}/ (2m),$ where $e$ and $m$ are the charge
and mass of the particle.  In Dirac's theory of charged point-like
spin-$1/2$ particles, $g=2$.  Quantum electrodynamics (QED) predicts
deviations from Dirac's value, as the charged particle can emit and
reabsorb virtual photons.  These QED effects slightly increase the
value of $g$. It is conventional to express the difference of $g$ from
2 in terms of the value of the so-called anomalous magnetic moment, a
dimensionless quantity defined as $a_l = (g - 2)/2$, with
$l = e, \mu$. Over the last decade, the muon magnetic dipole moment
has maintained a long-standing discrepancy of about $3.7\sigma$ between
the Standard Model (SM) prediction and the Brookhaven E821
experimental measurement~\cite{Jegerlehner:2009ry,Davier:2010nc,Aoyama:2020ynm}.  Very recently,
the Muon $g-2$ Experiment at Fermilab released its first results, which
in combination with the previous E821 measurement lead to a new
experimental average of the muon anomalous magnetic dipole moment of
$a_\mu^{\rm exp} = 116592061(41) \times
10^{-11}$~\cite{Abi:2021gix}. The difference
$\delta a_\mu \equiv a_\mu^{\rm exp} - a_\mu^{\rm SM} = (251 \pm 59) \times 10^{-11}$ has
a significance of $4.2\sigma$. In this paper we investigate whether
this discrepancy can be explained in the context of low-mass-scale
strings~\cite{Antoniadis:1998ig}.  Before proceeding, it is important
to stress that the absence of anomalous magnetic moment in a
supersymmetric abelian gauge theory~\cite{Ferrara:1974wb} does not
connect to the models discuss herein since the brane configuration is not supersymmetric. 

Our calculations are framed in the context of intersecting D-brane
models; namely, we consider extensions of the  SM based on open strings ending on D-branes, with gauge bosons due to
strings attached to stacks of D-branes and chiral matter due to
strings stretching between intersecting
D-branes~\cite{Blumenhagen:2000wh,Angelantonj:2000hi,Aldazabal:2000cn,Aldazabal:2000dg,Blumenhagen:2000ea,Ibanez:2001nd,Blumenhagen:2001te,Cvetic:2001tj,Cvetic:2001nr}.
Intersecting D-brane models encase a set of building block ground
rules, which can be used to assemble UV completions of the SM with the
effective low energy theory inherited from properties of the
overarching string theory.  For these models, the elemental unit of
gauge invariance is a $U(1)$ field, and therefore a stack of $N$
identical D-branes sequentially gives rise to a $U(N)$ theory with the
associated $U(N)$ gauge group. If many types of D-brane are present in
the model, the gauge group grows into product form $\prod U(N_i)$,
where $N_i$ specifies the number of D-branes in each stack. For $N =
2$, the gauge group can be $Sp(1) \cong SU(2)$ rather than
$U(2)$.\footnote{In the presence of orientifolds, one also obtains
  orthogonal and symplectic gauge groups.}  For further details, see
e.g.~\cite{Blumenhagen:2005mu,Blumenhagen:2006ci}.

The minimal embedding of the SM particle spectrum requires at least
three brane stacks~\cite{Antoniadis:2000ena} leading to three distinct
models of the type $U(3) \times U(2)\times U(1)$ that were
classified in~\cite{Antoniadis:2000ena, Antoniadis:2004dt}. Only one
of them (model C of~\cite{Antoniadis:2004dt}) has baryon number as a
gauge symmetry that guarantees proton stability (in perturbation
theory), and can be used in the framework of low mass scale string
compactifications. Besides, since the charge associated to the $U(1)$
of $U(2)$ does not participate in the hypercharge combination, $U(2)$
can be replaced by the symplectic $Sp(1)$ representation of
Weinberg-Salam $SU(2)_L$, leading to a model with one extra $U(1)$
added to the hypercharge~\cite{Berenstein:2006pk}. The SM embedding in
four D-brane stacks leads to many more models that have been
classified in~\cite{Antoniadis:2002qm,Anastasopoulos:2006da}.  Whether
low-mass-scale strings are realized in nature is yet to be answered,
and the search for new physics signals of intersecting D-brane models
is still one of the goals of current day
research~\cite{Anchordoqui:2007da,Anchordoqui:2008ac,Anchordoqui:2008hi,Lust:2008qc,Anchordoqui:2008di,Anchordoqui:2009mm,Anchordoqui:2009ja,Anchordoqui:2011ag,Anchordoqui:2011eg,Anchordoqui:2012wt,Anchordoqui:2012fq,Anchordoqui:2014wha,Anchordoqui:2015uea,Celis:2015eqs,Anchordoqui:2015jxc,Anchordoqui:2016rve,Anchordoqui:2010zs,Anchordoqui:2009bn,Anchordoqui:2012qu,Anchordoqui:2011nh,Anchordoqui:2016kmu,Anchordoqui:2020tlp}.

In D-brane models there are three types of contribution to the
anomalous magnetic moment of the muon: the one from anomalous massive
$U(1)$ gauge bosons, the one from excited string states (excitations
of $l^*$ are expected to appear in the $s$-channel, the poles occur at
$\sqrt{s} = \sqrt{n} \, M_s$, with
$n=1,\, 2,\, \dots$ and $M_s$ the string scale~\cite{Veneziano:1968yb}), and that of Kaluza-Klein
(KK) modes.  While the anomalous $U(1)$ Lagrangian permits a direct one
loop calculation of the anomalous magnetic
moment~\cite{Kiritsis:2002aj}, a direct calculation of the
contribution from Regge recurrences is not possible due to the
nonrenormalizability of the theory. However, it was conjectured
in~\cite{Kiritsis:2002aj} that the contributions from string
oscillators and KK states must be largely suppressed.  In
this paper we check thoroughly the various contributions to the
anomalous magnetic moment of the muon in three and four D-brane stacks
realizations of the SM. The layout of the paper is as follows.  In
Sec.~\ref{sec:2} we reconsider the contribution from anomalous $U(1)$
gauge bosons and  derive new constraints on the parameter space
imposed by recent LHC data. In Sec.~\ref{sec:3} we use sum rules
methods to calculate the contribution from Regge excitations, and we
verify the strong suppression of this contribution conjectured
in~\cite{Kiritsis:2002aj}.  The paper wraps up with some conclusions
presented in Sec.~\ref{sec:4}. Before proceeding we note that the
latest Fermilab data already lead to several new physics
interpretations with connections to other fundamental problems in
particle physics, astrophysics, and cosmology~\cite{Zhu:2021vlz,Yin:2021mls,Keung:2021rps,Athron:2021iuf,Aboubrahim:2021rwz,Bhattacharya:2021ggm,Kawamura:2021ygg,Baer:2021aax,Buras:2021btx}.

\section{Contributions from anomalous massive $\bm{U(1)}$ gauge bosons}
\label{sec:2}

To develop our program in the simplest way, we work within the construct of
minimal models with 3 and 4 stacks of D-branes.

\subsection{3 stack models}

For 3 stack models, the canonical gauge group is
$U(3) \times U(2) \times U(1)$, with stacks labeled $a$, $b$, and $c$,
respectively~\cite{Antoniadis:2000ena}. In the bosonic sector, the
open strings terminating on the QCD stack $a$ contain the standard
$SU(3)_C$ octet of gluons $g_\mu^a$ and an additional $U(1)_a$ gauge
boson $C_\mu$, most simply the manifestation of a gauged baryon number
symmetry: $U(3) \sim SU(3)_C\times U(1)_a$. On the $U(2)$ stack the
open strings correspond to the electroweak gauge bosons $A_\mu^a$, and
again an additional $U(1)_b$ gauge field $X_\mu$.  So the associated
gauge groups for these stacks are $SU(3)_C \times U(1)_a,$ $SU(2)_L
\times U(1)_b$, and $U(1)_c$, respectively.  The quantum numbers of quarks and leptons in each family are given by
\ba
&Q &({\bf 3},{\bf 2};1,1+2z,0)_{1/6}\nonumber\\
&u^c &({\bf\bar 3},{\bf 1};-1,0,0)_{-2/3}\nonumber\\
&d^c &({\bf\bar 3},{\bf 1};-1,0,1)_{1/3}\label{charges}\\
&L   &({\bf 1},{\bf 2};0,1,z)_{-1/2}\nonumber\\
&l^c &({\bf 1},{\bf 1};0,0,1)_1\nonumber
\ea
where $z=0,-1$. The charge assignments for the two Higgs doublets read
\ba
H\ \ ({\bf 1},{\bf 2};0,1+2z,1)_{1/2}\quad &H'\ \ ({\bf 1},{\bf 2};0,-(1+2z),0)_{1/2}
\label{Higgs}
\ea  
The relations for $U(N)$
unification, $g'_a = g_a/\sqrt{6}$ and $g'_b = g_b/2$, hold only at
$M_s$ because the $U(1)$ couplings ($g'_a$, $g'_b$, $g'_c$)
run differently from the non-Abelian $SU(3)$ ($g_a$) and $SU(2)$
($g_b$)~\cite{Anchordoqui:2011eg}.

We can perform a unitary transformation on the gauge fields $A_i
=U_{ij} \widetilde{A}_j $ (with $A_Y = \widetilde{A}_1,\ A_\xa =
\widetilde{A}_2,\ A_\xb = \widetilde{A}_3$) to diagonalize their mass
matrix. The $U(1)$ brane can be  independent of the other branes and
has in general a different gauge coupling $g_c$. In the model of~\cite{Antoniadis:2000ena},
however, the $U(1)$ brane was located on top of either the color or
the weak D-branes, and consequently $g_c$ being equal to either $g_a$
or $g_b$. Here we relax the additional constraint imposed in~\cite{Antoniadis:2000ena}  and instead set $g_a = y g_c$ at the string scale
$M_s$. As a result there are two free parameters $y, \xt$ in
$U_{ij}$. $A_\xa, A_\xb$ are the anomalous $U(1)$ gauge fields, whose
charges depend on $z$. For $z=0$, the anomaly free hypercharge is
given by
\be
Q_{Y} = Q_c-{Q_b\over 2} + {2Q_a\over 3}\,,
\ee
yielding
\ba
Q_{\xa}= -Q_c \frac {1} {\sqrt 2 y} \left(16 + 9 x^2\right) \sin \theta+\frac{\sqrt{2}}{\sqrt{3} x} Q_b \left(2 \cos \theta \sqrt{16+9 x^2+12 y^2} -3 \sqrt{3} x\, y \sin \theta \right) \nn
 +\frac 1 {\sqrt 6}Q_a \left(3 x \cos \theta \sqrt{16+9 x^2+12 y^2}+8 \sqrt{3} y \sin \theta\right)\,,\label{rotchargone}
\ea
and
\ba
Q_{\xb}= \frac {1}{\sqrt{2}y} Q_c \left(16 + 9 x^2\right) \cos \theta+\frac{\sqrt{2}}{\sqrt{3} x} Q_b \left(2 \sin \theta \sqrt{16+9 x^2+12 y^2}+3 \sqrt{3} x\, y \cos \theta\right) \nn
 +\frac 1 {\sqrt 6} Q_a \left(3 x \sin \theta \sqrt{16+9 x^2+12 y^2}-8 \sqrt{3} y \cos \theta\right)\,,
\ea
while for $z=-1$, the hypercharge is found to be
\begin{equation}
Q_{Y} = Q_c+{Q_b\over 2} + {2Q_a\over 3}\,,
\end{equation}
yielding
\ba
Q_{\xa}= -\frac {1}{\sqrt{2}y}  Q_c \left(16 + 9 x^2\right)  \sin \theta+\frac{\sqrt{2}}{\sqrt{3} x} Q_b \left(2 \cos \theta \sqrt{16+9 x^2+12 y^2} +3 \sqrt{3} x\, y \sin \theta \right) \nn
 +\frac 1 {\sqrt 6}Q_a \left(-3 x \cos \theta \sqrt{16+9 x^2+12 y^2}+8
   \sqrt{3} y \sin \theta\right) \label{rotchargone}
 \ea
 and
\ba
Q_{\xb}= \frac {1}{\sqrt{2}y} Q_c \left(16+9 x^2\right)  \cos \theta+\frac{\sqrt{2}}{\sqrt{3} x} Q_b \left(2 \sin \theta \sqrt{16+9 x^2+12 y^2}-3 \sqrt{3} x\, y \cos \theta\right) \nn
 -\frac 1 {\sqrt 6}Q_a \left(3 x \sin \theta \sqrt{16+9 x^2+12 y^2}+8 \sqrt{3} y \cos \theta\right)\,,
\ea
with
\begin{equation}
  x = \frac {g_a/\sqrt 3}{g_b/ \sqrt 2} \, .
\end{equation}  
Throughout we adopt the convention for which the $U(1)$ coupling is $g'_c = g_c /\sqrt 2$. The three coupling constants are related to $g_Y$ at the string scale $M_s$ by~\cite{Anchordoqui:2011eg} 
\begin{equation}
\frac{2}{g_c^2}+\frac{4 \left(\frac{1}{2}\right)^2}{g_b^2}+\frac{6
    \left(\frac{2}{3}\right)^2}{g_a^2}=\frac{1}{g_Y^2} \, .
\label{diegote}
\end{equation}
  
\begin{figure}[tbp] 
\begin{minipage}[t]{0.49\textwidth}
    \postscript{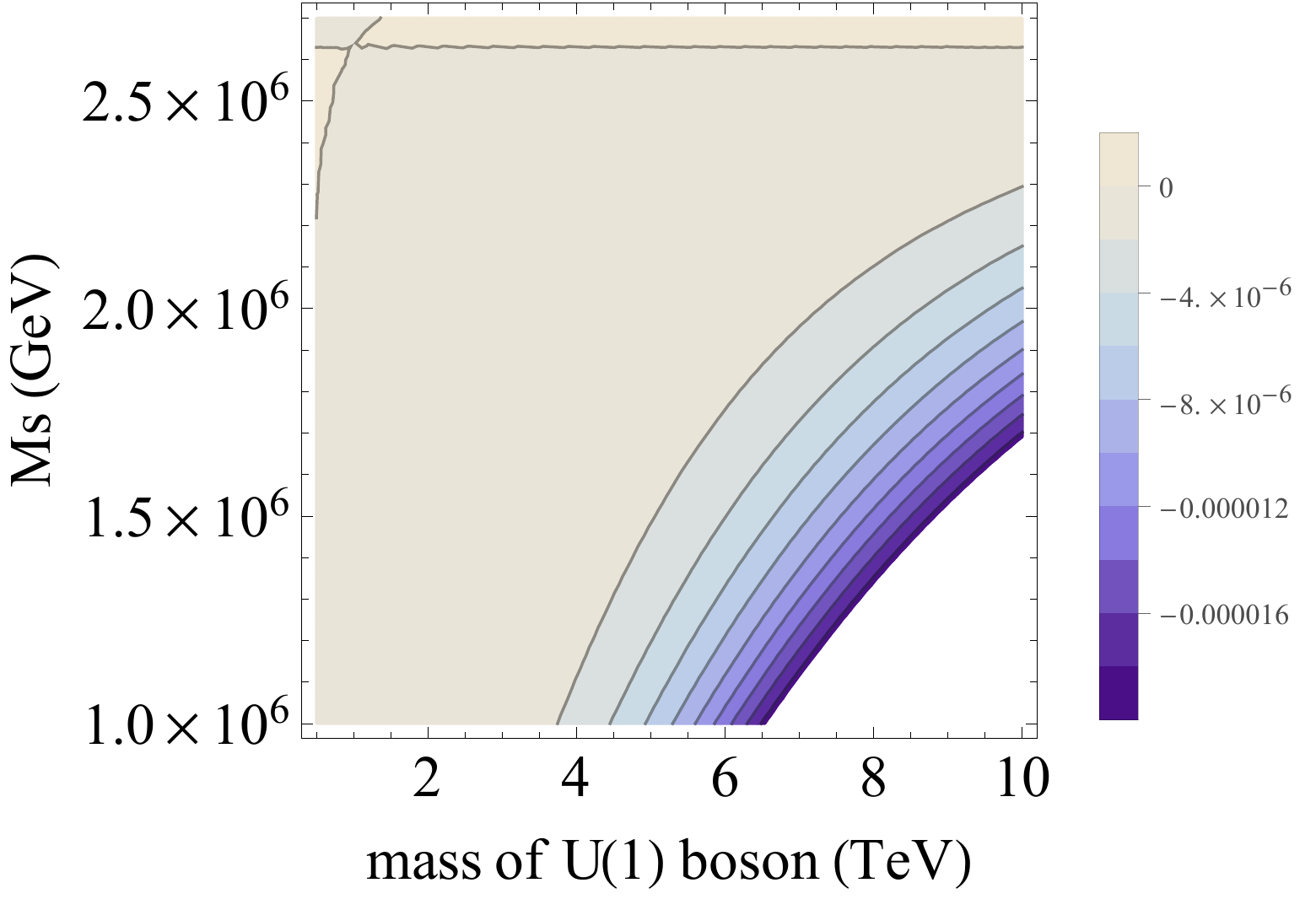}{0.99} 
\end{minipage} 
\hfill \begin{minipage}[t]{0.49\textwidth}
  \postscript{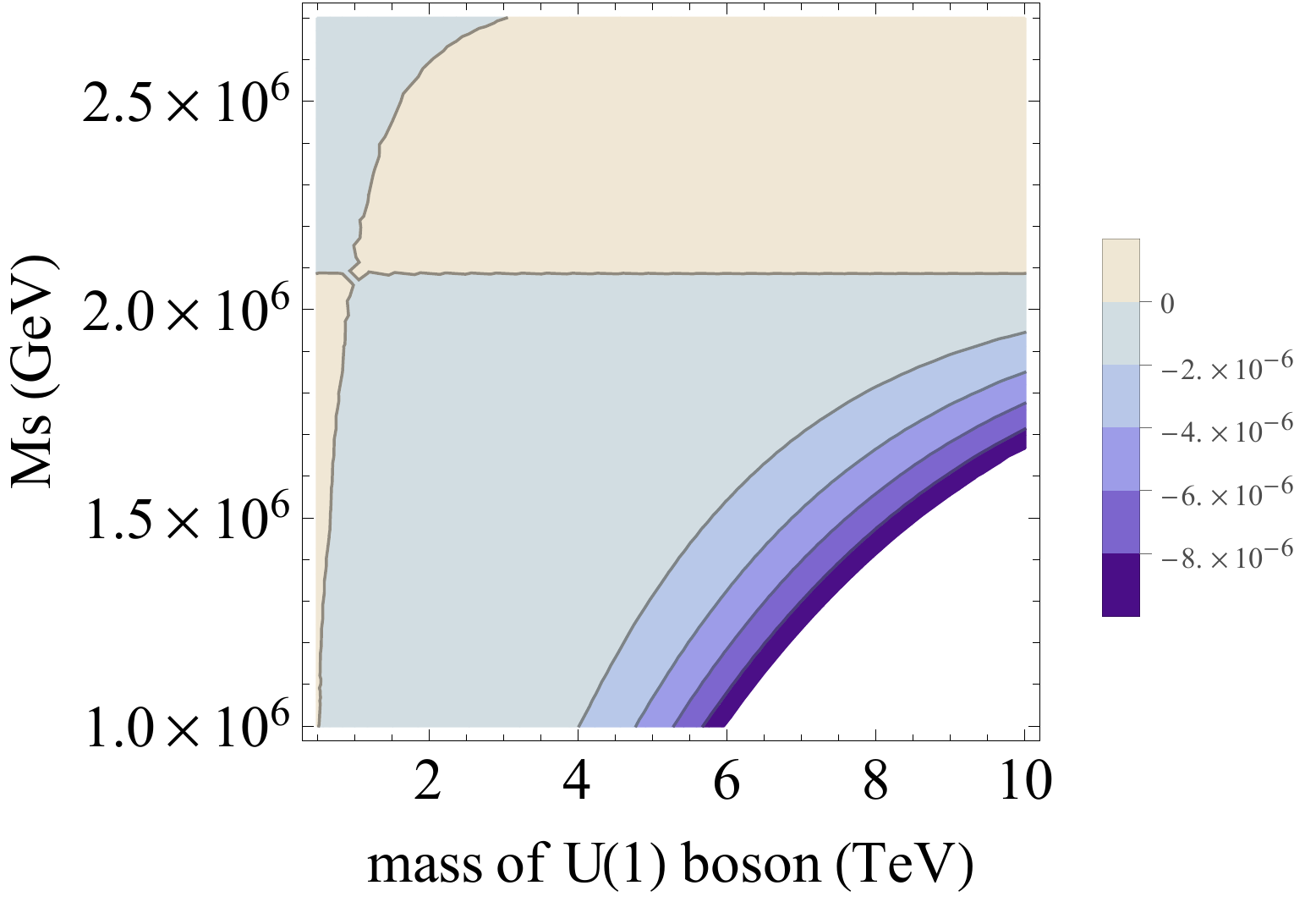}{0.99} 
\end{minipage} 
\caption{Allowed regions for 3-stack models. The left panel is for
  $z=0$ while the right panel is for $z=-1$. By setting $\mu_\xa =
  1$~TeV, we are left with two parameters $M_s$ and $\mu_\xb$
  (x-axis). The contours are for the values of the discriminant of
  equations \er{tanthetaFir} and \er{tanthetaSec} (more precisely we
  multiply the discriminant by $\mu_\xb^4$). The allowed regions are
  where the discriminant is positive and we can have some combination
  of $\mu_\xa$, $\mu_\xb$, $M_s$ so that the deviation of anomalous magnetic moment is from the contribution of the anomalous $U(1)$ bosons.}  
\label{fig:1} 
\end{figure}

The anomalous magnetic moment of the muon in the D-brane realization
of the SM, ${a}^{U(3)\times U(2) \times U(1)}_\mu$,
can be computed from one-loop correction to
the muon vertex. Generalizing the procedure described in~\cite{Kiritsis:2002aj}  for $y \neq 1$, we calculate the
contribution to ${a}^{U(3)\times U(2) \times U(1)}_\mu$ from the
anomalous $U(1)$ exchanged diagrams as well as the axion
diagrams. When these contributions are added
 to the SM prediction we obtain 
\be
{a}^{U(3)\times U(2) \times U(1)}_\mu={a}^{\rm SM}_\mu
+g^2 \sum_{i={\xa},\xb} {Q_{i L}^2-3Q_{i L}Q_{i R}+Q_{i R}^2\over 12\pi^2} 
\Big({m_l\over {\mu}_i}\Big)^2 +{h^2\over 16\pi^2}\,,
\label{AMM}
\ee
where $Q_{iL}$ and $Q_{iR}$ are the $U(1)$ charges for the left and
right-handed muons respectively, $g$ is the effective
coupling for the anomalous $U(1)$ gauge fields and its value is 
\be
g = \frac{g_c}{\sqrt{16+9 x^2} \sqrt{16+9 x^2+12 y^2} }\,,
\ee
$h$ is the Planck constant, and $\mu_{\alpha}, \mu_\beta$ are the masses
of the anomalous $U(1)$ fields. Substituting in $Q_\xa$ and $Q_\xb$, we get the following equation relating
$\mu_\xa$ and $\mu_\xb$ to the difference $\delta a_\mu$ of anomalous magnetic moment
of the muon for $z=0$,
\ba
g_3^2 m^2 \left\{t^2 \lsb 16 \mu_\xa^2 y^2 \left(9 x^2+12 y^2+16\right)+3 \mu_\xb^2 x^2 \left(-18 \left(9 x^2+16\right) y^2+\left(9 x^2+16\right)^2+36 y^4\right)\rsb \rc\nn
 +3 \mu_\xa^2 x^2 \left(-18 \left(9 x^2+16\right) y^2+\left(9 x^2+16\right)^2+36 y^4\right)+16 \mu_\xb^2 y^2 \left(9 x^2+12 y^2+16\right) \nn
\left. -12 t x y (\mu_\xa^2-\mu_\xb^2)\left(9 x^2-4 y^2+16\right) \sqrt{27 x^2+36 y^2+48}\right\}\nn
\label{tanthetaFir}-72 \pi ^2 \mu_\xa^2 \mu_\xb^2 ({\xd}{a_\mu}-{a_\mu}{}_{\phi'}) \left(t^2+1\right) x^2 \left(9 x^2+16\right) y^2 \left(9 x^2+12 y^2+16\right)=0,\quad
\ea
where $t = \tan \xt$ ($\xt$ is the angle that appears in $U_{ij}$) and ${a_\mu}{}_{\phi'}$ is the
contribution from the axion (which is proportional to $h$~\cite{Kiritsis:2002aj}). Likewise, for the $z=-1$ model we obtain
\ba
g_3^2 m^2 \left\{t^2 \left(16 \mu_\xa^2 y^2 \left(9 x^2+12 y^2+16\right)+3 \mu_\xb^2 x^2 \left(30 \left(9 x^2+16\right) y^2+5 \left(9 x^2+16\right)^2+36 y^4\right)\right)
\right. \nn
+3 \mu_\xa^2 x^2 \left(30 \left(9 x^2+16\right) y^2+5 \left(9 x^2+16\right)^2+36 y^4\right) +16 \mu_\xb^2 y^2 \left(9 x^2+12 y^2+16\right)\nn
\left. -4 t x y (\mu_\xa^2-\mu_\xb^2)\left(45 x^2+12 y^2+80\right) \sqrt{27 x^2+36 y^2+48}\right\}\nn
\label{tanthetaSec}-72 \pi ^2 \mu_\xa^2 \mu_\xb^2 ({\xd}{a_\mu}-{a_\mu}{}_{\phi'}) \left(t^2+1\right) x^2 \left(9 x^2+16\right) y^2 \left(9 x^2+12 y^2+16\right)=0.\quad
\ea
We note that $\tan \xt$ has to be real and therefore the discriminant of
\er{tanthetaFir} (or \er{tanthetaSec}) has to be positive
definite. The values of $x,y$ are determined by $g_b, g_c, g_Y$ at the
string scale $M_s$ and we have three free parameters $\mu_\xa$,
$\mu_\xb$, $M_s$. For simplicity, to analyze the $(\mu_\xb,M_s)$
parameter space we set $\mu_\xa = 1~{\rm TeV}$. The corresponding
allowed regions are given in  Fig.~\ref{fig:1}.

\subsection{4 stack models}

\begin{figure}[tbp]
\postscript{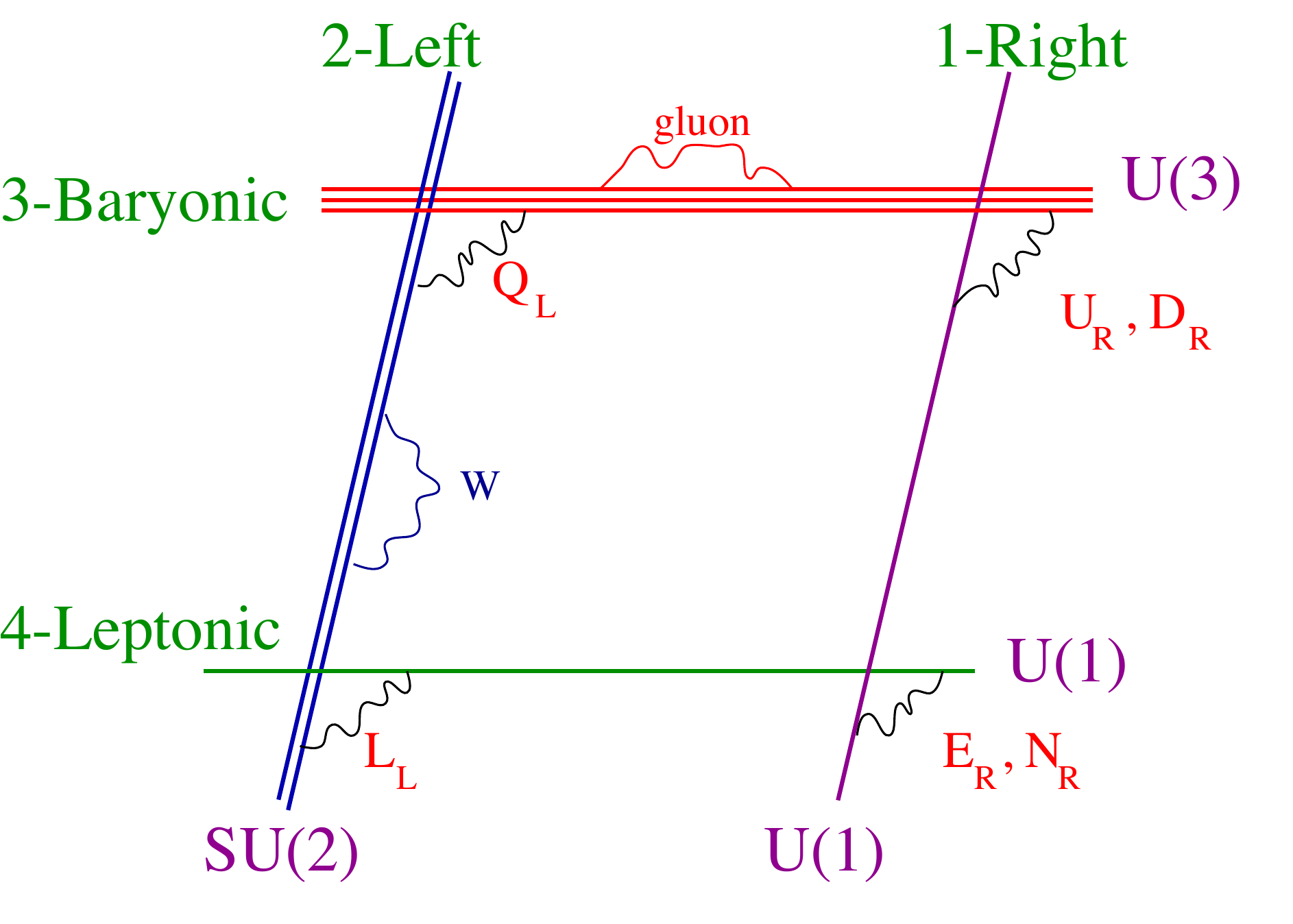}{0.8}
\caption{Pictorial representation of the $U(3)_B \times SU(2)_L \times
  U(1)_{L} \times U(1)_{I_R}$ D-brane model, for stacks $a,b,c,d$, respectively.}
\label{cartoon}
\end{figure}

\begin{table}
  \caption{Chiral  spectrum.}
\begin{center}
\begin{tabular}{ccccccc}
\hline
\hline
~~~Fields~~~ & ~~~Sector~~~  & ~~~Representation~~~ & ~~~$Q_B$~~~ & ~~~$Q_L$~~~ & ~~~$Q_{I_R}$~~~ & ~~~$Q_Y$~~~ \\
\hline
 $U_R$ &   $\phantom{^*}3 \leftrightharpoons 1^*$ &  $(3,1)$ & $-1$ & $\phantom{-}0$ & $- 1$ & $\phantom{-}\frac{2}{3}$  \\[1mm]
  $D_R$ & $3 \leftrightharpoons 1$  & $( 3,1)$&    $-1$ & $\phantom{-}0$
                                                                                & $\phantom{-}1$ & $-\frac{1}{3}$   \\[1mm]
  $L_L$ & $4 \leftrightharpoons 2$ &  $(1,2)$&    $\phantom{-}0$ &  $- 1$ & $\phantom{-}0$ & $-\frac{1}{2}$ \\[1mm]
  $E_R$ & $4 \leftrightharpoons 1$ &  $(1,1)$&   $\phantom{-}0$ & $-1$ &  $\phantom{-} 1$ & $- 1$ \\[1mm]
 $Q_L$ & $3 \leftrightharpoons 2$ &  $(3,2)$& $-1$ & $\phantom{-}0 $ & $\phantom{-} 0$ & $\phantom{-} \frac{1}{6}$    \\[1mm]
   $N_R$  &  $\phantom{^*}4 \leftrightharpoons 1^*$  &   $(1,1)$& $\phantom{-}0$ & $-1$ & 
$- 1$ & $\phantom{-} 0$ \\ [1mm]
$H$ & $2 \leftrightharpoons 1$ &  $(1,2)$ & $\phantom{-}0$ & $\phantom{-}0$ & $-1 $ &
$\phantom{-}\frac{1}{2}$  \\ [1mm]
$H''$ & $4 \leftrightharpoons 1$ &   $(1,1)$ & $\phantom{-}0$ & $\phantom{-}1$ & $\phantom{-}1$  & $\phantom{-}0$ \\ 
\hline
\hline
\label{table}
\end{tabular}
\end{center}
\end{table}

Now, let us consider  models with 4 stacks of
D-branes. If we consider next-to-minimal constructs where in
the $b$ stack we choose projections leading to the symplectic $Sp(1)$
representation of Weinberg-Salam, the
gauge
extended sector, $U(3)_B \times SU(2)_L \times U(1)_{L} \times
U(1)_{I_R}$, has two additional $U(1)$ symmetries.
A schematic representation of the
D-brane construct is shown in Fig.~\ref{cartoon} and  the quantum numbers of the chiral spectrum are summarized in Table~\ref{table}.

\begin{figure}[tbp] 
  \postscript{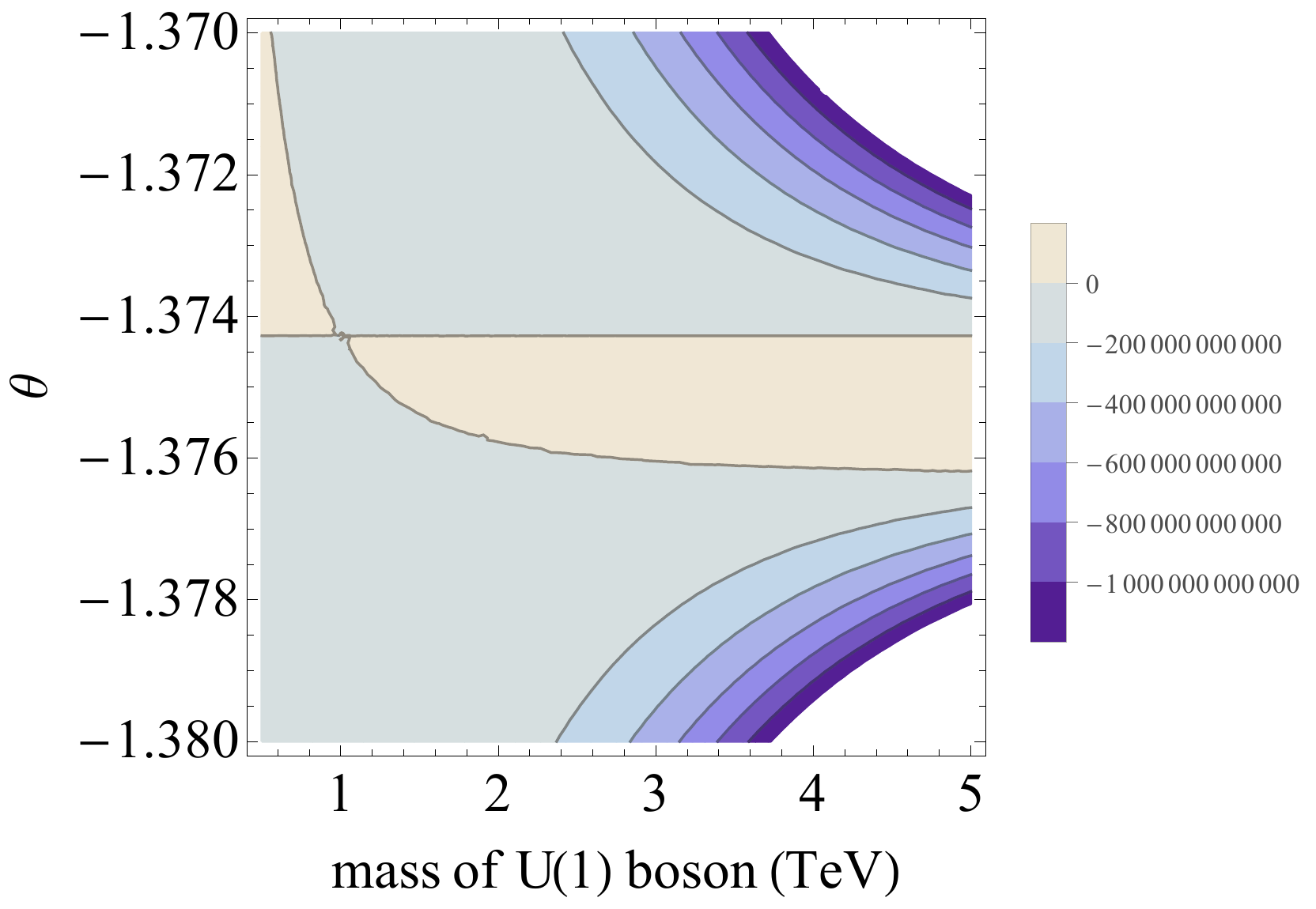}{0.99} 
\caption{Allowed regions for 4-stack model. By setting $\mu_\xa = 1$TeV and $M_s = 2 \times 10^3$, we are left with two parameters $\mu_\xb$ (x-axis) and $\xt$ (y-axis). The contours are again for the values of the discriminant of a quadratic equation set up by requiring measured anomalous magnetic moment given by \er{AMM}. As we can see there are regions with positive discriminant where we can find appropriate $\phi$ so that the $U(1)$ anomalous bosons can account for the deviation of anomalous magnetic moment.}  
\label{fig:2} 
\end{figure}

The resulting $U(1)$ content gauges the baryon number $B$ [with
$U(1)_B \subset U(3)_B$], the lepton number $L$, and a third
additional abelian charge $I_R$ which acts as the third isospin
component of an $SU(2)_R$.  Contact with gauge structures at TeV
energies is achieved by a field rotation to couple diagonally to
hypercharge $Y_\mu$. Two of the Euler angles are determined by this
rotation, and the hypercharge is given by
\be
Q_{Y} = \frac {Q_c} 2 - {Q_a\over 6} - {Q_d\over 2}\,,
\ee
The charges $Q_\xa$ and $Q_\xb$ of the two anomalous $U(1)$ are given by 
\ba
Q_{\xa}& =& \frac 1 {\sqrt 2} g_c Q_c \cos \theta \sin \phi+\frac 1 {\sqrt 2} g_d Q_d (\sin \theta \sin \psi \sin \phi+\cos \psi \cos \phi )\nn
& & +\frac{1}{\sqrt{6}} g_a Q_a (\sin \theta \cos \psi \sin \phi-\sin \psi \cos \phi),\label{rotchargone3}
\ea
\ba
Q_{\xb}& =& \frac 1 {\sqrt 2} g_c Q_c \cos \theta\cos \phi+\frac{1}{\sqrt{2}}g_d Q_d (\sin \theta\sin \psi \cos \phi-\cos \psi \sin \phi)\nn
& & +\frac{1}{\sqrt{6}}g_a Q_a (\sin \theta\cos \psi \cos \phi+\sin \psi \sin \phi),\label{rotchargone2}
\ea
where $g_c, g_d$ are the coupling constants of the two $U(1)$ branes
and we again include the $1/\sqrt 2$ in the coupling. The gauge couplings are related to $g_Y$ by~\cite{Anchordoqui:2011eg} 
\be
\frac{2 \left(\frac{1}{2}\right)^2}{g_c^2}+\frac{2
  \left(\frac{1}{2}\right)^2}{g_d^2}+\frac{6
  \left(\frac{1}{6}\right)^2}{g_a^2}=\frac{1}{g_Y^2} \, .
\ee
As previously noted, the Euler angles $\xt,\psi,\phi$ parameterize the
unitary transformation $A_i =U_{ij} \widetilde{A}_j $. We can
parametrize $\psi$ in terms of the $g_d$ coupling constant and $\theta$,
\be
\psi \to -\text{arcsin}\lb -\frac{\sqrt{2} g_Y}{2 g_d \cos \theta
} \rb \, .
\ee
For later convenience, we can also parameterize the coupling $g_c$ using the angle $\xt$ 
\be
g_c \to -\frac{\sqrt{2} g_Y}{2 \sin \theta} \, .
\ee
The anomalous magnetic moment is again given by \er{AMM}, which leads
to a quadratic equation of $t=\tan \phi$. To analyze the $(\mu_\xb,M_s)$
parameter space we set $\mu_\xa = 1~{\rm TeV}$. Now we have one more
parameter ($\xt$) than before and following the same philosophy the
string scale $M_s$ is picked to be $2 \times 10^3$ TeV. The allowed
regions of the parameter space, parameterized by $\xt, \mu_\xb$, are
encapsulated in Fig.~\ref{fig:2}.

\subsection{LHC constraints}

We now turn to confront our findings with null results from LHC
searches of new heavy vector bosons~\cite{Sirunyan:2021khd,Sirunyan:2019vgj}. To this end we duplicate
the procedure outlined above but scanning on $\mu_{\xa}$. The possible highest mass scale for $\mu_{\xa}$ is in general around
 1~TeV. For 3 stack models, it is 0.6~TeV for $z = 0$, and 1.4~TeV for
 $ z = -1$. The 4-stack model does not help much as the upper bound of
 the mass scale is also around 1.5~TeV. The dominant constraint comes from the
 requirement that neither $U(1)$ coupling is above $2 \pi$: $g_c$
 becomes ${\cal O} (1)$ at the string scale; its running is between
 the anomalous $U(1)$ mass and the string scale. The
 4-stack model has some extra freedom in that the string scale can
 vary (in the 3-stack case the string scale is fixed by the
 coupling). Since the couplings of the anomalous $U(1)$ bosons are in
 general much larger than the SM coupling (especially when their masses
 are large), we can compare the constraints on $\mu_\xa$ with LHC
 limits on the generalized sequential model~\cite{Accomando:2010fz}, containing the
 sequential SM boson gauge boson that has SM-like couplings to SM
 fermions~\cite{Altarelli:1989ff}. We can see that the discrepancy
 $\delta a_\mu$ reported by Fermilab cannot be  accommodated within
 the $1\sigma$ contours without being in
tension with LHC data~\cite{Sirunyan:2021khd,Sirunyan:2019vgj}, unless
the massive $U(1)$ gauge boson $Z'$ is
leptophilic~\cite{Buras:2021btx}. 

We consider
the resonant production cross section of $\sigma(pp\to Z' \to \ell
\ell)$. Under the narrow width approximation, the cross section can be written in the form of $c_u w_u + c_d w_d$, where $w_u, w_d$ are given by model-independent parton distribution functions~\cite{Accomando:2010fz}. The coupling of $Z'$ with up and down quarks (assuming same coupling to three families)
are encoded in $c_u,c_d$. More precisely, for a generic coupling between $Z'$ and fermion $f$ 
\begin{equation}
Z^\prime_{\mu}\gamma^{\mu}(\bar f_L\epsilon_L^f f_L +\bar f_R \epsilon_R^f
f_R)\,,
\end{equation}
the coefficients $c_u$ and $c_d$ take the following form
\begin{equation}
c_{f}=({\epsilon_L^f}^2+{\epsilon_R^f}^2)\textrm{Br}(\ell^+\ell^-)\,.
\end{equation}
We compute the branching faction $\textrm{Br}(\ell^+\ell^-)$ by including
only the decay channels to leptons and quarks. The total decay rate is given
by
\begin{equation}
\Gamma_{Z'}=
\frac{1}{24\pi}M_{Z'}\left[
9\sum_{q=u,d} ({\epsilon_L^q}^2+{\epsilon_R^q}^2) +3\sum_{\ell=e,\nu} ({\epsilon_L^\ell}^2+{\epsilon_R^\ell}^2)
\right].
\end{equation}
Because of the constraint
(\ref{diegote}),  there are two free parameters (for a given
string scale $M_s$): $\phi$ and $g_d (M_s)$. Setting the mass of $Z'$ to 2~TeV, we then search
over the parameter space to get the smallest possible values of $c_u,
c_d$. For simplicity, the combination of $\sqrt{c_u^2+c_d^2}$ is
considered. We find that the optimal value of $\phi$ generally
suppresses the couplings to left-handed quarks and the remaining
couplings to the right-handed quarks are controlled by $g_d$. In the
best case scenario, $g_c(M_s)$ is set to $2\pi$ at $M_s = 10~{\rm
  TeV}$, the corresponding cross section (or rather $\sqrt{c_u^2+c_d^2}
\sim 8.4\times 10^{-5}$) is roughly 2 percent of that given by the
sequential standard model boson, saturating the
LHC limit~\cite{Sirunyan:2021khd}. We note that the
branching fraction to leptons is close to $1$ due to the small
coupling to quarks. The signal can be further reduced by including
other decay channels. Moreover, the largest possible $g_c(M_s)$ also
gives the most contribution to $a_\mu$.

From (\ref{AMM}) we see that each anomalous $U(1)$ in the
4-stack model provides the following contribution to $a_\mu$
\begin{equation}
a_\mu^{(i)} = -\frac {Q_{iL}^2 - 3 Q_{iL} Q_{iR} + Q_{iR}^2}{12 \pi^2}
\left( \frac{m_l} {\mu_i} \right)^2,\quad i = \alpha, \beta\, .
\end{equation}
We can choose $A_\alpha$ to be the
leptophilic $Z'$ and plug into (\ref{rotchargone3}) the charges from
Table~\ref{table}. The $L$ and $R$ charges then read 
\begin{equation}
  Q_{\alpha L} = -\frac{g_c \cos \theta \sin \phi}{\sqrt{2}} 
\end{equation}
and
\begin{equation}
  Q_{\alpha R} = \frac{ g_d (\sin \theta \sin \phi \sin \psi + \cos
    \phi \cos \psi )-g_c \cos \theta \sin \phi}{\sqrt{2}} \, .
\end{equation}
Such a $Z'$ boson gives $a_\mu^{(\alpha)} = 9.9 \times 10^{-11}$,
which is still not enough to explain the observed
discrepancy. $A_\beta$ shall be much heavier to avoid the LHC bound
and its contribution is negligible. If we
require $\mu_\alpha \agt 5~{\rm TeV}$ and $\mu_\beta \agt 5~{\rm
  TeV}$, then $\delta a_\mu \alt 3 \times 10^{-11}$, with $g_c \sim 1$.

\section{Contributions from excited string states}
\label{sec:3}

The low energy limit of the spin-flip forward Compton amplitude in the
laboratory frame of the target lepton $l$ is related to the square
of its anomalous magnetic moment. The assumption of analyticity and
sufficient convergence permits an unsubtracted dispersion relation for
this amplitude and, together with the optical theorem, a sum rule for
$a_l^2$~\cite{Gerasimov:1965et,Drell:1966jv}, where 
\begin{equation}
  (a_l^{\rm QED}+ a_l^{\rm non-QED})^2 = \frac{m_l^2} {2 \pi^2\alpha}\
  \int_{s_{\rm th}}^\infty \frac{ds}{s} \ \Delta\sigma \  ,
\label{rhs}
\end{equation}
$\alpha$ is the QED fine-structure constant, $s$ is the square
ceneter-of-mass energy, and where in
obvious notation~\cite{Jaffe:1987hv}
\begin{equation}
\Delta \sigma = \frac{1}{2} \left[ \left( \sigma_{1,1/2} -
    \sigma_{-1,1/2} \right) + \left( \sigma_{1,-1/2} -
    \sigma_{-1,-1/2} \right) \right] \,  .
\end{equation}
The first of the 3 terms on the LHS of the equation is canceled on the RHS by the integral 
containing only $\Delta\sigma_{\rm QED}$, so that the truncated sum rule is~\cite{Brodsky:1980zm,Goldberg:1981gc,Goldberg:1999gc} 
\begin{equation}
 2\; ( a_{\rm QED} )  \ ( a_{\rm non-QED}) \  +  \  (a_{\rm non-QED})\
 ^2 = \frac{m_l^2}{2\pi^2\alpha} \ \int_{s_{\rm th}}^\infty \frac{ds}{s} \
 \Delta\sigma_{\rm non-QED} \, .
\label{rhs1}
\end{equation}
With the use of collinear string amplitudes~\cite{Anchordoqui:2008hi}, a straightforward calculation
shows that the contribution of a single spin 1/2 or 3/2 resonance of
mass $M_s$ to the right-hand side of~(\ref{rhs1}) is given by
\begin{equation}
\Delta \sigma_{\rm non-QED} = 2 \pi e^2 \, \delta (s - M_s^2) \,,
\end{equation}
which gives a contribution $4 (m_l/ M_s)^2$. However, in the string
spectrum, there is the possibility of cancelation between the
different spin contributions to the RHS of (\ref{rhs}).

Further insights into the problem are as follows: the tree level contribution to
$\Delta \sigma_{\rm non-QED}$ from right handed muons is proportional to
\begin{equation}
\sigma_{ \xg_L \mu_R} - \sigma_{\xg_R \mu_R } \equiv \sum_n \lb \sigma_{\xg_L
\mu_R }^n - \sigma_{\xg_R \mu_R }^n \rb,
\end{equation}
where $\sigma_{\xg_L \mu_R }^n$ is the total cross section for
outgoing $\mu_R^*$ Regge excited states at level $n$. Note that the
muon has its momentum along $+z$ direction. For $n=1$, the right hand
side can be simplified to
\begin{equation}
\sigma_{B_L \mu_R }^1 - \sigma_{B_R \mu_R }^1 \propto \lb  \left|F_{B_L \mu_R }^{1,J = 3/2}\right|^2 - \left|F_{B_R \mu_R }^{1,J = 1/2}\right|^2 \rb \,,
\end{equation}
where $F_{B_{L,R} \mu_{L,R} }^{n,J}$ are the collinear amplitudes of
the $\mu_R^*$ Regge excitation of spin $J$ (to simplify notation,
gauge group indices have been omitted) and level $n$. If $\xD \xs =
0$, then the relation
\begin{equation}
F_{B_R \mu_R }^{1,J = 1/2} = F_{B_L \mu_R }^{1,J = 3/2}
\label{mainrel}
\end{equation}
should hold. Using~\cite{Anchordoqui:2008hi}
\begin{equation}
F_{+\frac 1 2+1\alpha_3a_4}^{\alpha J=1/2} =   F_{-\frac 1 2-1\alpha_3a_4}^{\alpha J=1/2} = F_{+\frac 1 2-1\alpha_3a_4}^{\alpha J=3/2} = F_{-\frac 1 2+1\alpha_3a_4}^{\alpha J=3/2} = \sqrt{2}\, g\, M  \, T^{a_4}_{\alpha_3\,\alpha} 
\label{colq}
\end{equation}
it is straightforward to see that (\ref{mainrel}) is satisfied. The
amplitude for $\mu_L \xg \to \mu_L^*$ is proportional to a linear
combination of $F_{X_L \mu_L}^{n,J}$ and $F_{A^3_L \mu_L }^{n,J}$. More
explicitly, we have
\begin{equation}
\sigma_{\xg_L \mu_L }^1 \propto \left|\eta \ C_W \ F_{X_L \mu_L }^{n,J = 1/2} + S_W \ F_{A^3_L \mu_L }^{n,J = 1/2} \right|^2 \,,
\end{equation}
where $C_W \equiv \cos \theta_W$, $S_W = \sin \theta_W$, and
$\theta_W$ is the Weinberg angle. The cross section
$\sigma_{\xg_R \mu_L }^1$ can be expressed in a similar form (with
$J=1/2$ replaced by $J=3/2$). Then $\sigma_{\xg_R \mu_L }^1 -
\sigma_{\xg_L \mu_L }^1 = 0$ follows from Eq.~\eqref{colq}.

For $n > 1$, it is actually more convenient to show the cancelations
in terms of the helicity amplitudes, i.e., $A_{-1,-1/2} =
A_{+1,-1/2}$. These amplitudes are given by~\cite{Anchordoqui:2008hi},
\begin{equation}
A_{+1,-1/2} \sim {\cal M}_{\mu_L(1) X_R(2) \to \mu_L(3) X_R(4)}  = 
g'_b{}^2 
\frac{\langle 14\rangle^2}{\langle 12\rangle\langle 23\rangle} \lb \frac s t V_s + \frac u t V_u \rb \label{ggqql}
\end{equation}
and 
\begin{equation}
 A_{-1,-1/2} \sim {\cal M}_{\mu_L(1) X_L(2) \to \mu_L(3) X_L(4)}  = 
g'_b{}^2 
\frac{\langle 12\rangle^2}{\langle 14\rangle\langle 34\rangle} \lb \frac s t V_s + \frac u t V_u \rb  \, . \label{ggql} 
\end{equation}
The amplitude ${\cal M}_{\mu_L A^3 \to \mu_L A^3}$ can be obtained by replacing $g'_b{}^2$ by $g_b^2$. Note that we restore the form factors $V_u, V_s$, which gives all the pole terms. They are defined as
$V_t =V(s,t,u) ,$ $V_u=V(t,u,s) ,$  $V_s=V(u,s,t),$ with
\begin{equation}
V(s,t,u)= {\Gamma(1-s/M_s^2)\ \Gamma(1-u/M_s^2)\over
    \Gamma(1+t/M_s^2)}\ .
    \end{equation}
For forward scattering,
we have the photon momenta $k_4 = - k_2$ and it is not difficult to see that the tree level contribution to $\Delta
\sigma_{\rm non-QED}$ gets canceled, i.e., $A_{-1,-1/2}
= A_{+1,-1/2}$. 

The cancelation may be violated at order $\alpha$ by differing mass
shifts for different spins. Support for this is manifest by the
different widths (imaginary parts of the bubble) for $J=1/2$, $J=3/2$~\cite{Anchordoqui:2014wha},
which will imply that the mass shifts will differ.  In passing we note
that resonance production accompanied by single photon emission from
the electron line, $\gamma  l \to \gamma l^*$, does not contribute to
$\Delta \sigma_{\rm non-QED}$~\cite{Brodsky:1995fj}.

We now turn to the explicit mass shift calculation. The Breit-Wigner
form for unstable particles $\xs^{J=1/2, 3/2}$ is given by
\be \label{crosssection}\xs(E) = \frac {2J+1}{(2S_1+1)(2S_2+1)} \frac
{4\pi} {k^2} \lsb \frac {\xG^2/4}{(E-E_0)^2 + \xG^2 /4} \rsb B_{\rm
  in} B_{\rm out}, \ee where $B_{\rm in} = \xG_{\rm in}/\xG$ and
$B_{\rm out} = \xG_{\rm in}/\xG$. The out states need to be integrated
and therefore $B_{\rm out} =1$. Since $\xG_{\rm in}^{J=1/2} = 2
\xG_{\rm in}^{J=3/2}$ we have $(2J+1)\xG_{\rm in}^{J = 3/2}=
(2J+1)\xG_{\rm in}^{J = 1/2}$.  We denote the energy of the photon in
the rest frame of the electron by $\nu$, and it is related to $s$ by
$(\nu + m_l)^2 - \nu^2 = s \equiv E^2$. The integration over $d\nu $
then gives \ba \frac {m_l^2}{2\pi^2 \xa}\int \frac {d\nu} \nu \xs(E) &
= & \frac {m_l^2}{2\pi^2 \xa} \frac {16\pi(2J+1)\xG_{in}}{3} \int
\frac {2E d E}{E^2-m_l^2} \frac 1 {\xG E^2} \lsb \frac
{\xG^2/4}{(E-M_s)^2 + \xG^2 /4} \rsb \nn & = & \frac {64 m_l^2
  (2J+1)\xG_{in}} {3\xa \xG ^3} \int \frac {d x} {\pi\lb x^2 -
  x_1^2\rb x} \lsb \frac {1}{(x-x_0)^2 + 1} \rsb\,, \label{penco}\ea
where $x = 2E/\xG$. To avoid the infrared divergence at $x = x_1$ we
set a cutoff and integrate over the energy range $(M_s - 2 \xG, M_s +
2 \xG)$, that is $(x_0 - 4, x_0 +4)$.  The integral in the second line
of (\ref{penco}) is found to be \be \int_{-4}^{+4} \frac {d x} {\pi\lb
  x^2 - x_1^2\rb x} \lsb \frac {1}{(x-x_0)^2 + 1} \rsb = \frac{2 \tan
  ^{-1}(4)}{\pi }\epsilon ^3-\frac{12 \left(\tan
    ^{-1}(4)-4\right)}{\pi } \epsilon ^5 + \CO(\xe^7)\,, \ee where
$\xe = 1 /x_0 = \xG/( 2M_s)$. The difference between cross sections
into $J=1/2$ and $J=3/2$ reads, \be \frac {m_l^2}{2\pi^2 \xa}\int
\frac {d\nu} \nu \xD \xs \sim \frac {2 m_l^2 \times 2\xG_{\rm in}^{J=
    1/ 2}} {3\xa M_s^3}\frac{12 \left(4-\tan ^{-1}4\right)}{\pi
}\lb\frac {\xD \xG^2} {M_s^2}\rb \,.
\label{diecisiete}
\ee For 3-stack models, the $J=1/2$ decay width of the left-handed
leptons is \be \xG^{J=1/2}_L = \frac {1} {8 \times 1} \frac {{g'_c}^2}
{4\pi} M_s + \frac {2^2-1} {8 \times 2} \frac {{g_b}^2} {4\pi} M_s +
\frac {1} {8 \times 2} \frac {{g_b'}^2} {4\pi} M_s\,, \ee while for
right-handed leptons we have \be \xG^{J= 1/2}_R = \frac {2^2} {8
  \times 1} \frac {{g'_c}^2} {4\pi} M_s\, .  \ee For 4-stack models,
we obtain \be \xG^{J=1/2}_L = \frac {1} {8 \times 1} \frac {{g'_c}^2}
{4\pi} M_s + \frac {2^2-1} {8 \times 2} \frac {{g_b}^2} {4\pi} M_s \,,
\ee and \be \el{decaywidthL} \xG^{J= 1/2}_R = \frac {1} {8 \times 1}
\frac {{g'_d}^2} {4\pi} M_s + \frac {1} {8 \times 1} \frac {{g'_c}^2}
{4\pi} M_s\, .  \ee Before proceeding we note again that $\xG^{J= 3/
  2}_{L,R} = \frac 1 2 \xG^{J= 1/ 2}_{L,R}$.

The sum rule applies for either the left or right handed leptons. The
correction to $a_l^2$ (from mass shift) are proportional to the
difference in the square of widths of left- and right-handed excited
leptons respectively. This implies that the magnetic moment is not
proportional to the spin.

It is convenient to parametrize the RHS of (\ref{diecisiete}) as
follows \be \frac {m_l^2} {M_s^2}\frac {4\xG_{in}(J=\frac 1 2)} {3\xa
  M_s}\frac{12 \left(4-\tan ^{-1}4\right)}{\pi }\lb\frac {\xD \xG^2}
{M_s^2}\rb = 4 \left(\frac{\alpha}{\pi}\right) \, \lb \frac {m_l^2}
{M_s^2} \rb \ \kappa_{L,R} \, .  \ee Using the expressions for the
total decay widths given above and $\xG_{\rm in}^{J= 1 /2} = \xa
M_s/8$~\cite{Anchordoqui:2008hi} we evaluate (\ref{diecisiete}) to
obtain $\kappa_{L,R}$. For 3-stack models, $\kappa_L = 0.0064$
and $\kappa_R = 0.00022$. For 4-stack models, the ratio $g'_c/g'_d$
is driven by $g_Y$~\cite{Anchordoqui:2014wha}. By requiring both $g'_c$ and $g'_d$ to be less than 1,
we constrain the couplings to be within $(0.183, 1)$. Note that
$\kappa_{L,R}$ and the constraint are invariant under a swap of $g'_c$
and $g'_d$. In this coupling range, the interval for kappa becomes $0.0063
<\kappa_L < 0.037$ while $0.00023 < \kappa_R < 0.015$.

Now we would like to estimate the contribution from higher resonance
states, $n>1$. For simplicity, we consider the case in which the photon
and the muon have parallel spins. The amplitude of two gauge bosons
and two fermions is given by \be
\CM(g^{-}_1,f_2^-,g^{+}_3,\bar f_4^+)~=~ 2\, g^2\,
\delta^{\beta_4}_{\beta_2}\,{\langle 12\rangle^2\over \langle
  23\rangle\langle 34\rangle}\
\Big[(T^{a_1}T^{a_3})^{\alpha_2}_{\alpha_4}\ {s\over t}\ V_s+
(T^{a_3}T^{a_1})^{\alpha_2}_{\alpha_4}\ {u\over t}\ V_u\Big]\ , \ee
where $g$  is the $U(N)$ coupling constant, $\langle ij \rangle$ are
the standard spinor products written in the notation of~\cite{Mangano:1990by,Dixon:1996wi,Taylor:2017sph},
and the $U(N)$ generators are normalized according to ${\rm Tr}\, (T^a
T^b) = \delta^{ab}/2$~\cite{Lust:2008qc}. The function $ V_u$ has poles at $s = n M_s^2$,
\begin{equation}
V_{u}(n)=V(s,u,t)\approx\frac{1}{s-nM_s^{2}}\times
\frac{M_s^{2-2n}}{(n-1)!}\prod_{J=0}^{n-1}(t+M_s^{2}J)\,,\label{Vexpansion}
\end{equation}
and near the poles the amplitude takes the following form
\begin{equation}
\CM(g^{-}_1,f_2^-,g^{+}_3,\bar f_4^+) \to 
2\, g^2\, (T^{a_3}T^{a_1})^{\alpha_2}_{\alpha_4}\,\delta^{\beta_4}_{\beta_2}\,\frac{M_s^{2}}{s-nM_s^{2}}\sum_{j=0}^{n-1}
c^{(n)}_j
d^{j+\frac 1 2}_{-\frac 1 2, -\frac 1 2}(\xt)\,.
\end{equation}
The partial width of the $n$th resonance with angular momentum $J$ decaying
into $g+f$ with parallel spins reads
\begin{equation}
\label{partialw}
\Gamma(R_{n,J}\to g^\pm
f^\pm)=g^2\delta\,\frac{c_{j}^{(n)}M_s^2}{16(2J+1)\pi \sqrt n M_s} \, .
\end{equation}
To compute the coefficients $c_k^{(n)}$, we first write the resonance amplitude
as
\ba
\CM(g^{-}_1,f_2^-,g^{+}_3,\bar f_4^+) &\to &
2\, g^2\, (T^{a_3}T^{a_1})^{\alpha_2}_{\alpha_4}\,\delta^{\beta_4}_{\beta_2}\,\sqrt{-\frac s u}\  {u\over t}\ \hat V_u\ \nn
& \equiv & \frac{M_s^{2}}{s-nM_s^{2}} \sqrt{-\frac u s} f(-x,n) \,,
\ea
where
\begin{align}
\label{fxn}
f(x,n) & =\Big[\frac{n}{(n-1)!}\Big]\underbrace{[\frac{nx}{2}-(\frac{n}{2}-1)][\frac{nx}{2}-(\frac{n}{2}-2)]
\cdots[\frac{nx}{2}+(\frac{n}{2}-2)][\frac{nx}{2}+(\frac{n}{2}-1)]}_{n-1\; \zt{factors}}\nonumber \\
 & =\Big[\frac{n}{(n-1)!}\Big]\left(\frac{nx}{2}-\frac{n}{2}+1\right)_{(n-1)}.
\end{align}
The Pochhammer Symbol is defined as follows,
\begin{equation}
(x)_{n}=\frac{\Gamma(x+n)}{\Gamma(x)}=x(x+1)...(x+n-1)=\sum_{k=0}^{n}(-1)^{n-k}s(n,k)x^{k},\label{Stirling1}
\end{equation}
where $s(n,k)$ is the Stirling number of the first kind.
For odd $n$, we obtain\footnote{The computation for even $n$ is
  similar.}
\begin{align}
f(x,n) & =\Big[\frac{n}{(n-1)!}\Big]\left(\frac{nx}{2}-\frac{n}{2}+1\right)_{(n-1)}\nonumber \\
 &
 =\frac{n}{(n-1)!}(a_{n-1}^{(n)}x^{n-1}+a_{n-3}^{(n)}x^{n-3}+\cdots+a_{2}^{(n)}x^{2}+a_{0}^{(n)})\, ,
\end{align}
with
\begin{equation}
a_{k}=\sum_{i=0}^{n-1-k}\frac{(-1)^{n-1-k}}{2^{k+i}}s(n-1,k+i)\left(\begin{array}{c}
k+i\\
i
\end{array}\right)(n)^{k}(n-2)^{i}.
\end{equation}
It is convenient to rewrite $\sqrt{-\frac u s}f(-x,n)$ as
\begin{align}
&\sqrt{-\frac u s}f(-x,n) \nn 
& =\frac{n}{(n-1)!}\lb \frac{x+1} 2\rb^{\frac 1 2}(a_{n-1}^{(n)}(-x)^{n-1}+a_{n-3}^{(n)}(-x)^{n-3}+\cdots+a_{2}^{(n)}(-x)^{2}+a_{0}^{(n)})\nonumber \\
& =\lb \frac{x+1} 2\rb^{-\frac 1 2} \lb c_{n}^{(n)}P_{n}^{(0,-1)}+c_{n-1}^{(n)}P_{n-1}^{(0,-1)}+\cdots+c_{1}^{(n)}P_{1}^{(0,-1)}\rb\,.\label{expandf}
\end{align}
where in the last line we have used $d^{j+\frac 1 2}_{-\frac 1 2, -\frac 1 2} (\xt) = \lb \cos \frac
\xt 2 \rb^{-1} P^{(0,-1)}_j(\cos \xt)$, with $P^{(\xa,\xb)}_j(x)$  the Jacobi Polynomials. Note that $(x+1)  x^m/2$ can be expanded in terms of the Jacobi Polynomials,
\begin{equation}
\left( \frac{x+1} 2\right) x^{m}=\sum_{i=m+1,m,\cdots}d_i{}^m P_{i}^{(0,-1)}(x),
\end{equation}
where 
\begin{align}
d_i{}^m & =2i \int^{+1}_{-1} x^m P_{i}^{(0,-1)}(x) dx \nn
& = i\left(\frac{2^{\frac {i-m} 2} \left((-1)^i+(-1)^m\right) m!}{\lb \frac {m-i} 2\rb! \left(m+i+1\right)!!}+\frac{2^{\frac {i-m-1} 2}\left((-1)^i-(-1)^m\right) m!}{\left(\frac{m-i+1}{2}\right)!\left(m+i\right)!!}\right)\,.
\end{align}
We can then obtain $c_{j}^{(n)}$ as
\begin{align}
c_{j}^{(n)} & =\frac{n}{(n-1)!}\sum_{k=0}^{\frac{n-j}{2}}a_{j-1+2k}d^{j-1+2k}_j\,. \end{align}
Using \er{fxn}, \er{Stirling1}, and \er{expandf} the sum of $c_{j}^{(n)}$ can
be obtained
\be
\sum_j c_{j}^{(n)} = \frac {f(1,n)}{P_j^{(0,-1)}(1)} = n, 
\ee
where we also used the fact that $P_j^{(\xa,\xb)}(1) = C^{n+\xa}_n$. As a
result, the partial width \er{partialw} satisfies the following relation
\be
\sum_J \frac {(2J+1)\xG_{\rm in}} {\sqrt n M_s} = g^2\delta\,\frac{1}{16 \pi }\,.
\ee
We note that this $n$-independent quantity appears in the Breit-Wigner
form\er{crosssection}. In the narrow width approximation, it is the factor multiplying
the delta function $\xd (s - n M_s^2)$. In this case, we can see that the contribution to
$a_l^2$ from level $n$ is proportional to $n^{-1}$. In other words, particles
of masses $M$ together contribute $\sim (m_l/M)^2$ to $a_l^2$. 
Assuming that the ratio of the total decay width and mass
\be
\frac {\xG_{\zt{tot}}} M = \frac {g^2}{4\pi} \frac \CC 4\,,
\ee
remains a constant, it is straightforward to see that the same behavior applies
for a Breit-Wigner form.

\begin{table}[htp]
\caption{Lower limits on $M_s/{\rm TeV}$.}
\begin{center}
\begin{tabular}{ccc}
\hline
\hline
~~~~~~~~~~~~~~~~chirality~~~~~~~~~~~~~~~~ &  ~~~~~~~~~~~~~~~~3-stack~~~~~~~~~~~~~~~~& ~~~~~~~~~~~~~~~~4-stack~~~~~~~~~~~~~~~~  \\
\hline
$L$   & $0.66$ & $0.66$ to $1.6$    \\
$R$  & $0.12$ & $0.13$ to  $1.0$    \\
\hline
\hline
\end{tabular}
\end{center}
\label{boundMs}
\end{table}%

With $a_l^{\rm QED} \simeq \alpha/(2\pi)$, we find on solving
Eq.~(\ref{rhs1})
\begin{equation}
a_l^{\rm non-QED}    \simeq 4 \left( \frac{m_l}{M} \right)^2\  \kappa \, b(n_0)\,,
\end{equation}
where $b(n_0)$ denotes the total contribution from levels $n <
n_0$. As we discussed elsewhere~\cite{Anchordoqui:2014wha},  for $n_0 \agt 40$, the total decay width
grows bigger than the spacing of mass levels and the resonance picture breaks
down. Specializing now to the muon, we see that $a_\mu = 1.16 \times
10^{-3}$ and \mbox{$a_\mu^{\rm non-QED}  < 3 \times
10^{-9}$~\cite{Abi:2021gix}}  translates to a lower limit on the
string scale, that is:
\be
M_s > 4\  {\rm TeV}\; \sqrt{\kappa \ b(n_0)} \, .
\ee
The bounds for 3- and 4-stack models summarized in
Table~\ref{boundMs}  show that the contributions from string
excitations to the anomalous magnetic moment of the muon are largely
suppressed, as conjectured in~\cite{Kiritsis:2002aj}.

\section{Conclusions}
\label{sec:4}

Very recently, the Fermilab Muon $g - 2$ experiment released its first
measurement of the anomalous magnetic moment, which is in full
agreement with the previous measurement at Brookhaven and pushes the
world average deviation from the SM expectation $\delta a_\mu$ to a
significance of $4.2\sigma$~\cite{Abi:2021gix}.  Galvanized by this brand new result we
have reexamined the contributions to $\delta a_\mu$ from anomalous
$U(1)$ gauge bosons and excitations of the string. We have shown that
while the contribution from Regge recurrences is strongly suppressed,
the contribution from the heavy vector bosons can help ameliorate
(though not fully eliminate) the $\delta a_\mu$ discrepancy.

In closing, we note that KK winding modes could provide a non-negligible 
contribution to $\delta a_\mu$. It was conjectured in~\cite{Kiritsis:2002aj} that the KK
contribution would be also suppressed. However, we argue herein that
this speculation is model dependent, because it is based on the statement that the
compactification scale is of order the string scale. An order of
magnitude estimate can be obtained by using the truncated sum rule: setting $a_{\rm QED} =
\alpha/(2\pi)$ in (\ref{rhs1}) we obtain 
\begin{equation}
 ( a_{\rm non-QED}) \  + \frac{\pi}{\alpha}  \  (a_{\rm non-QED})\
 ^2 = \frac{m_l^2}{2\pi \alpha^2} \ \int_{s_{\rm th}}^\infty \frac{ds}{s} \
 \Delta\sigma_{\rm non-QED} \, .
\label{kk1}
\end{equation}
If $a_{\rm non-QED} \ll \alpha/\pi$, then the left-hand-side of
(\ref{kk1}) is dominated by its first term,
\begin{equation}
 ( a_{\rm non-QED}) \  \sim \frac{m_l^2}{2\pi \alpha^2} \ \int_{s_{\rm
     th}}^\infty \frac{ds}{s} \
 \Delta\sigma_{\rm non-QED} \, .
\label{kk2}
\end{equation}
Assuming that both the left- and right-hand-side of (\ref{kk2}) can be
written as power laws in $\alpha$ (or loop number), on the
left-hand-side we expect  $a_{\rm non-QED}^{\rm KK}$ to be $\cal {O} (\alpha)$, as seen
from triangle diagrams containing heavy stuff.  Then,  on the
right-hand-side 
$\Delta \sigma_{\rm non-QED}^{\rm KK}$ (for excitation $M_j$) needs to be of  ${\cal O}(\alpha^3/M_j^2)$, which yields 
\begin{equation}
a_{\rm non-QED}^{\rm KK} \sim \sum_j \ \left(\frac{m_l}{M_j}\right)^2 \alpha \, .
\label{kk3}
\end{equation}
As we have shown in Fig.~\ref{fig:1}, if $\mu_\xa
\sim 1~{\rm TeV}$, for particular choice of gauge couplings, we can bridge the $\delta a_\mu$ gap
reported by the Muon $g - 2$ Collaboration. However, LHC experiments have set a 95\% CL bound
$\mu_\xa < 5~{\rm TeV}$~\cite{Sirunyan:2021khd,Sirunyan:2019vgj}.  We
can infer from (\ref{kk3}) that if the $U(1)$ retains the
same gauge couplings but $\mu_a \sim 5~{\rm TeV}$, then the contribution of each KK excitation will be suppressed by about
two orders of magnitude. Summation over all the KK modes may allow
recovering the suppression factor~\cite{Antoniadis:1993jp}. Note that the KK of other gauge
bosons with different couplings (e.g. hypercharge) would also
contribute. In addition, it is important to stress that {\it for the 4
  stack model exhibited in Fig.~\ref{cartoon}, the  KK excitations of
  the $U(1)$ which
lives in the lepton brane do
not have tree level couplings to hadrons and therefore their production
at the LHC would be suppressed, but these excitations could still yield the dominant contribution to $\delta
a_\mu$.} The future muon smasher will provide the final verdict on such
leptophilic KK excitations~\cite{Ali:2021xlw}. A direct string calculation of the KK contribution is under way and will be presented elsewhere.

\acknowledgments{This paper is dedicated to
  the memory of our dear colleague and sincere friend Haim Goldberg. The work of L.A.A. is supported by the by the
  U.S. National Science Foundation (NSF Grant PHY-1620661) and the
  National Aeronautics and Space Administration (NASA Grant
  80NSSC18K0464). The research of I.A. was partially performed as
  International professor of the Francqui Foundation, Belgium. The
  work of D.L. is supported by the Origins Excellence Cluster. The
  work of T.R.T is supported by NSF under Grant Number PHY–1913328.
  Any opinions, findings, and conclusions or recommendations expressed
  in this material are those of the authors and do not necessarily
  reflect the views of the NSF or NASA.}


\begin{thebibliography}{30}




  
\bibitem{Jegerlehner:2009ry} 
  F.~Jegerlehner and A.~Nyffeler,
  {\color{rossoCP3}  The muon $g-2$},
  Phys.\ Rept.\  {\bf 477}, 1 (2009)
  [arXiv:0902.3360 [hep-ph]].

\bibitem{Davier:2010nc}
M.~Davier, A.~Hoecker, B.~Malaescu and Z.~Zhang,
 {\color{rossoCP3}  Reevaluation of the hadronic contributions to the muon $g-2$ and to $\alpha(M^2_Z)$},
Eur. Phys. J. C \textbf{71} (2011), 1515
[erratum: Eur. Phys. J. C \textbf{72} (2012), 1874]
doi:10.1140/epjc/s10052-012-1874-8
[arXiv:1010.4180 [hep-ph]].


  
\bibitem{Aoyama:2020ynm}
T.~Aoyama \textit{et al.},
 {\color{rossoCP3} The anomalous magnetic moment of the muon in the Standard Model},
Phys. Rept. \textbf{887} (2020), 1-166
doi:10.1016/j.physrep.2020.07.006
[arXiv:2006.04822 [hep-ph]].


  
\bibitem{Abi:2021gix}
B.~Abi \textit{et al.} [Muon $g-2$ Collaboration],
  {\color{rossoCP3}   Measurement of the Positive Muon Anomalous Magnetic Moment to 0.46 ppm},
Phys. Rev. Lett. \textbf{126} (2021), 141801
doi:10.1103/PhysRevLett.126.141801
[arXiv:2104.03281 [hep-ex]].



  
\bibitem{Antoniadis:1998ig} 
  I.~Antoniadis, N.~Arkani-Hamed, S.~Dimopoulos and G.~R.~Dvali,
    {\color{rossoCP3} New dimensions at a millimeter to a Fermi and superstrings at a TeV},
  Phys.\ Lett.\ B {\bf 436}, 257 (1998)
  [hep-ph/9804398].


\bibitem{Ferrara:1974wb} 
  S.~Ferrara and E.~Remiddi,
    {\color{rossoCP3} Absence of the anomalous magnetic moment in a supersymmetric Abelian gauge theory},
  Phys.\ Lett.\ B {\bf 53}, 347 (1974).




\bibitem{Blumenhagen:2000wh} 
  R.~Blumenhagen, L.~Goerlich, B.~Kors and D.~L\"ust,
  {\color{rossoCP3} Noncommutative compactifications of type I strings on tori with magnetic background flux},
  JHEP {\bf 0010}, 006 (2000)
  doi:10.1088/1126-6708/2000/10/006
  [hep-th/0007024].



\bibitem{Angelantonj:2000hi} 
  C.~Angelantonj, I.~Antoniadis, E.~Dudas and A.~Sagnotti,
   {\color{rossoCP3} Type I strings on magnetized orbifolds and brane transmutation},
  Phys.\ Lett.\ B {\bf 489}, 223 (2000)
  doi:10.1016/S0370-2693(00)00907-2
  [hep-th/0007090].



\bibitem{Aldazabal:2000cn} 
  G.~Aldazabal, S.~Franco, L.~E.~Ibanez, R.~Rabadan and A.~M.~Uranga,
   {\color{rossoCP3} Intersecting brane worlds},
  JHEP {\bf 0102}, 047 (2001)
  doi:10.1088/1126-6708/2001/02/047
  [hep-ph/0011132].


\bibitem{Aldazabal:2000dg} 
  G.~Aldazabal, S.~Franco, L.~E.~Ibanez, R.~Rabadan and A.~M.~Uranga,
   {\color{rossoCP3} D = 4 chiral string compactifications from intersecting branes},
  J.\ Math.\ Phys.\  {\bf 42}, 3103 (2001)
  doi:10.1063/1.1376157
  [hep-th/0011073].


\bibitem{Blumenhagen:2000ea} 
  R.~Blumenhagen, B.~Kors and D.~L\"ust,
   {\color{rossoCP3} Type I strings with F flux and B flux},
  JHEP {\bf 0102}, 030 (2001)
  doi:10.1088/1126-6708/2001/02/030
  [hep-th/0012156].



\bibitem{Ibanez:2001nd} 
  L.~E.~Ibanez, F.~Marchesano and R.~Rabadan,
   {\color{rossoCP3} Getting just the standard model at intersecting branes},
  JHEP {\bf 0111}, 002 (2001)
  doi:10.1088/1126-6708/2001/11/002
  [hep-th/0105155].



\bibitem{Blumenhagen:2001te} 
  R.~Blumenhagen, B.~Kors, D.~L\"ust and T.~Ott,
   {\color{rossoCP3} The standard model from stable intersecting brane world orbifolds},
  Nucl.\ Phys.\ B {\bf 616}, 3 (2001)
  doi:10.1016/S0550-3213(01)00423-0
  [hep-th/0107138].


\bibitem{Cvetic:2001tj} 
  M.~Cvetic, G.~Shiu and A.~M.~Uranga,
   {\color{rossoCP3} Three family supersymmetric standard-like models from intersecting brane worlds},
  Phys.\ Rev.\ Lett.\  {\bf 87}, 201801 (2001)
  doi:10.1103/PhysRevLett.87.201801
  [hep-th/0107143].


\bibitem{Cvetic:2001nr} 
  M.~Cvetic, G.~Shiu and A.~M.~Uranga,
 {\color{rossoCP3} Chiral four-dimensional N=1 supersymmetric type 2A orientifolds from intersecting D6 branes},
  Nucl.\ Phys.\ B {\bf 615}, 3 (2001)
  doi:10.1016/S0550-3213(01)00427-8
  [hep-th/0107166].


\bibitem{Blumenhagen:2005mu} 
  R.~Blumenhagen, M.~Cvetic, P.~Langacker and G.~Shiu,
   {\color{rossoCP3} Toward realistic intersecting D-brane models},
  Ann.\ Rev.\ Nucl.\ Part.\ Sci.\  {\bf 55}, 71 (2005)
  doi:10.1146/annurev.nucl.55.090704.151541
  [hep-th/0502005].


\bibitem{Blumenhagen:2006ci} 
  R.~Blumenhagen, B.~Kors, D.~L\"ust and S.~Stieberger,
   {\color{rossoCP3} Four-dimensional string compactifications with D-branes, orientifolds and fluxes},
  Phys.\ Rept.\  {\bf 445}, 1 (2007)
  doi:10.1016/j.physrep.2007.04.003
  [hep-th/0610327].



\bibitem{Antoniadis:2000ena} 
  I.~Antoniadis, E.~Kiritsis and T.~N.~Tomaras,
   {\color{rossoCP3} A D-brane alternative to unification},
  Phys.\ Lett.\ B {\bf 486}, 186 (2000)
  doi:10.1016/S0370-2693(00)00733-4
  [hep-ph/0004214].

\bibitem{Antoniadis:2004dt} 
  I.~Antoniadis and S.~Dimopoulos,
   {\color{rossoCP3} Splitting supersymmetry in string theory},
  Nucl.\ Phys.\ B {\bf 715}, 120 (2005)
  doi:10.1016/j.nuclphysb.2005.03.005
  [hep-th/0411032].


\bibitem{Berenstein:2006pk} 
  D.~Berenstein and S.~Pinansky,
   {\color{rossoCP3} The minimal quiver standard model},
  Phys.\ Rev.\ D {\bf 75}, 095009 (2007)
  doi:10.1103/PhysRevD.75.095009
  [hep-th/0610104].

\bibitem{Antoniadis:2002qm} 
  I.~Antoniadis, E.~Kiritsis, J.~Rizos and T.~N.~Tomaras,
   {\color{rossoCP3} D-branes and the standard model},
  Nucl.\ Phys.\ B {\bf 660}, 81 (2003)
  doi:10.1016/S0550-3213(03)00256-6
  [hep-th/0210263].

\bibitem{Anastasopoulos:2006da} 
  P.~Anastasopoulos, T.~P.~T.~Dijkstra, E.~Kiritsis and A.~N.~Schellekens,
   {\color{rossoCP3} Orientifolds, hypercharge embeddings and the standard model},
  Nucl.\ Phys.\ B {\bf 759}, 83 (2006)
  doi:10.1016/j.nuclphysb.2006.10.013
  [hep-th/0605226].


\bibitem{Anchordoqui:2007da} 
  L.~A.~Anchordoqui, H.~Goldberg, S.~Nawata and T.~R.~Taylor,
  {\color{rossoCP3}  Jet signals for low mass strings at the LHC},
  Phys.\ Rev.\ Lett.\  {\bf 100}, 171603 (2008)
  doi:10.1103/PhysRevLett.100.171603
  [arXiv:0712.0386 [hep-ph]].


\bibitem{Anchordoqui:2008ac} 
  L.~A.~Anchordoqui, H.~Goldberg, S.~Nawata and T.~R.~Taylor,
  {\color{rossoCP3} Direct photons as probes of low mass strings at the CERN LHC},
  Phys.\ Rev.\ D {\bf 78}, 016005 (2008)
  doi:10.1103/PhysRevD.78.016005
  [arXiv:0804.2013 [hep-ph]].


\bibitem{Anchordoqui:2008hi} 
  L.~A.~Anchordoqui, H.~Goldberg and T.~R.~Taylor,
  {\color{rossoCP3} Decay widths of lowest massive Regge excitations of open strings},
  Phys.\ Lett.\ B {\bf 668}, 373 (2008)
  doi:10.1016/j.physletb.2008.09.003
  [arXiv:0806.3420 [hep-ph]].


\bibitem{Lust:2008qc} 
  D.~L\"ust, S.~Stieberger and T.~R.~Taylor,
  {\color{rossoCP3} The LHC string hunter's companion},
  Nucl.\ Phys.\ B {\bf 808}, 1 (2009)
  doi:10.1016/j.nuclphysb.2008.09.012
  [arXiv:0807.3333 [hep-th]].



\bibitem{Anchordoqui:2008di} 
  L.~A.~Anchordoqui, H.~Goldberg, D.~L\"ust, S.~Nawata, S.~Stieberger and T.~R.~Taylor,
  {\color{rossoCP3} Dijet signals for low mass strings at the LHC},
  Phys.\ Rev.\ Lett.\  {\bf 101}, 241803 (2008)
  doi:10.1103/PhysRevLett.101.241803
  [arXiv:0808.0497 [hep-ph]].


\bibitem{Anchordoqui:2009mm} 
  L.~A.~Anchordoqui, H.~Goldberg, D.~L\"ust, S.~Nawata, S.~Stieberger and T.~R.~Taylor,
  {\color{rossoCP3} LHC phenomenology for string hunters},
  Nucl.\ Phys.\ B {\bf 821}, 181 (2009)
  doi:10.1016/j.nuclphysb.2009.06.023
  [arXiv:0904.3547 [hep-ph]].


\bibitem{Anchordoqui:2009ja} 
  L.~A.~Anchordoqui, H.~Goldberg, D.~L\"ust, S.~Stieberger and T.~R.~Taylor,
  {\color{rossoCP3} String phenomenology at the LHC},
  Mod.\ Phys.\ Lett.\ A {\bf 24}, 2481 (2009)
  doi:10.1142/S021773230903196X
  [arXiv:0909.2216 [hep-ph]].


\bibitem{Anchordoqui:2009bn} 
  L.~A.~Anchordoqui, H.~Goldberg, D.~Hooper, D.~Marfatia and T.~R.~Taylor,
 {\color{rossoCP3}   Neutralino dark matter annihilation to monoenergetic gamma rays as a signal of low mass superstrings},
  Phys.\ Lett.\ B {\bf 683}, 321 (2010)
  doi:10.1016/j.physletb.2009.12.037
  [arXiv:0912.0517 [hep-ph]].




\bibitem{Anchordoqui:2010zs} L.~A.~Anchordoqui, W.~Z.~Feng,
  H.~Goldberg, X.~Huang and T.~R.~Taylor, {\color{rossoCP3} Searching
    for string resonances in $e^+e^-$ and $\gamma \gamma$ collisions},
  Phys.\ Rev.\ D {\bf 83}, 106006 (2011)
  doi:10.1103/PhysRevD.83.106006 [arXiv:1012.3466 [hep-ph]].



\bibitem{Anchordoqui:2011ag} 
  L.~A.~Anchordoqui, H.~Goldberg, X.~Huang, D.~L\"ust and T.~R.~Taylor,
  {\color{rossoCP3} Stringy origin of Tevatron $Wjj$ anomaly},
  Phys.\ Lett.\ B {\bf 701}, 224 (2011)
  doi:10.1016/j.physletb.2011.05.049
  [arXiv:1104.2302 [hep-ph]].


\bibitem{Anchordoqui:2011eg} 
  L.~A.~Anchordoqui, I.~Antoniadis, H.~Goldberg, X.~Huang, D.~L\"ust and T.~R.~Taylor,
  {\color{rossoCP3} $Z'$-gauge bosons as harbingers of low mass strings},
  Phys.\ Rev.\ D {\bf 85}, 086003 (2012)
  doi:10.1103/PhysRevD.85.086003
  [arXiv:1107.4309 [hep-ph]].


\bibitem{Anchordoqui:2011nh} 
  L.~A.~Anchordoqui and H.~Goldberg,
  {\color{rossoCP3} Neutrino cosmology after WMAP 7-year data and LHC first $Z'$ bounds},
  Phys.\ Rev.\ Lett.\  {\bf 108}, 081805 (2012)
  doi:10.1103/PhysRevLett.108.081805
  [arXiv:1111.7264 [hep-ph]].



\bibitem{Anchordoqui:2012wt} 
  L.~A.~Anchordoqui, I.~Antoniadis, H.~Goldberg, X.~Huang, D.~L\"ust, T.~R.~Taylor and B.~Vlcek,
  {\color{rossoCP3} LHC phenomenology and cosmology of string-inspired intersecting D-brane models},
  Phys.\ Rev.\ D {\bf 86}, 066004 (2012)
  doi:10.1103/PhysRevD.86.066004
  [arXiv:1206.2537 [hep-ph]].



\bibitem{Anchordoqui:2012fq} 
  L.~A.~Anchordoqui, I.~Antoniadis, H.~Goldberg, X.~Huang, D.~L\"ust, T.~R.~Taylor and B.~Vlcek,
  {\color{rossoCP3}  Vacuum stability of standard model$^{++}$},
  JHEP {\bf 1302}, 074 (2013)
  doi:10.1007/JHEP02(2013)074
  [arXiv:1208.2821 [hep-ph]].



\bibitem{Anchordoqui:2012qu} L.~A.~Anchordoqui, H.~Goldberg and
  G.~Steigman, {\color{rossoCP3} Right-handed neutrinos as the dark
    radiation: status and forecasts for the LHC}, Phys.\ Lett.\ B {\bf
    718}, 1162 (2013) doi:10.1016/j.physletb.2012.12.019
  [arXiv:1211.0186 [hep-ph]].





\bibitem{Anchordoqui:2014wha} 
L.~A.~Anchordoqui, I.~Antoniadis, D.~C.~Dai, W.~Z.~Feng, H.~Goldberg,
X.~Huang, D.~L\"ust, D.~Stojkovic,  and T.~R.~Taylor,
  {\color{rossoCP3}  String resonances at hadron colliders},
  Phys.\ Rev.\ D {\bf 90}, no. 6, 066013 (2014)
  doi:10.1103/PhysRevD.90.066013
  [arXiv:1407.8120 [hep-ph]].


\bibitem{Anchordoqui:2015uea} 
  L.~A.~Anchordoqui, I.~Antoniadis, H.~Goldberg, X.~Huang, D.~L\"ust and T.~R.~Taylor,
  {\color{rossoCP3} Stringy origin of diboson and dijet excesses at the LHC},
  Phys.\ Lett.\ B {\bf 749}, 484 (2015)
  doi:10.1016/j.physletb.2015.08.033
  [arXiv:1507.05299 [hep-ph]].

\bibitem{Celis:2015eqs}
A.~Celis, W.~Z.~Feng and D.~L\"ust,
 {\color{rossoCP3}  Stringy explanation of $b \rightarrow{} s \ell^+ \ell^-$ anomalies},
JHEP \textbf{02} (2016), 007
doi:10.1007/JHEP02(2016)007
[arXiv:1512.02218 [hep-ph]].



  
\bibitem{Anchordoqui:2015jxc} 
  L.~A.~Anchordoqui, I.~Antoniadis, H.~Goldberg, X.~Huang, D.~L\"ust and T.~R.~Taylor,
  {\color{rossoCP3} 750 GeV diphotons from closed string states},
  Phys.\ Lett.\ B {\bf 755}, 312 (2016)
  doi:10.1016/j.physletb.2016.02.024
  [arXiv:1512.08502 [hep-ph]].


\bibitem{Anchordoqui:2016rve} 
  L.~A.~Anchordoqui, I.~Antoniadis, H.~Goldberg, X.~Huang, D.~L\"ust and T.~R.~Taylor,
  {\color{rossoCP3} Update on 750 GeV diphotons from closed string states},
  Phys.\ Lett.\ B {\bf 759}, 223 (2016)
  doi:10.1016/j.physletb.2016.05.063
  [arXiv:1603.08294 [hep-ph]].




\bibitem{Anchordoqui:2016kmu} 
  L.~A.~Anchordoqui, I.~Antoniadis, H.~Goldberg, X.~Huang, D.~L\"ust and T.~R.~Taylor,
   {\color{rossoCP3} Minimal left-right symmetric intersecting D-brane model},
  Phys.\ Rev.\ D {\bf 95}, no. 2, 026011 (2017)
  doi:10.1103/PhysRevD.95.026011
  [arXiv:1611.09785 [hep-ph]].

\bibitem{Anchordoqui:2020tlp}
L.~A.~Anchordoqui, I.~Antoniadis, K.~Benakli and D.~L\"ust,
  {\color{rossoCP3} Anomalous $U(1)$ gauge bosons as light dark matter in string theory},
Phys. Lett. B \textbf{810} (2020), 135838
doi:10.1016/j.physletb.2020.135838
[arXiv:2007.11697 [hep-th]].


\bibitem{Veneziano:1968yb}
  G.~Veneziano,
  {\color{rossoCP3} Construction of a crossing - symmetric, Regge behaved amplitude for
  linearly rising trajectories},
  Nuovo Cim.\  A {\bf 57}, 190 (1968).

\bibitem{Kiritsis:2002aj} 
  E.~Kiritsis and P.~Anastasopoulos,
   {\color{rossoCP3} The anomalous magnetic moment of the muon in the D-brane realization of the standard model},
  JHEP {\bf 0205}, 054 (2002)
  [hep-ph/0201295].

\bibitem{Zhu:2021vlz}
B.~Zhu and X.~Liu,
 {\color{rossoCP3} Probing light dark matter with scalar mediator: muon $(g-2)$ deviation, the proton radius puzzle},
[arXiv:2104.03238 [hep-ph]].

\bibitem{Yin:2021mls}
W.~Yin,
 {\color{rossoCP3} Muon $g-2$ anomaly in anomaly mediation},
[arXiv:2104.03259 [hep-ph]].

\bibitem{Keung:2021rps}
W.~Y.~Keung, D.~Marfatia and P.~Y.~Tseng,
 {\color{rossoCP3} Axion-like particles, two-Higgs-doublet models, leptoquarks, and the electron and muon g-2},
[arXiv:2104.03341 [hep-ph]].


\bibitem{Athron:2021iuf}
P.~Athron, C.~Bal\'azs, D.~H.~Jacob, W.~Kotlarski, D.~St\"ockinger and H.~St\"ockinger-Kim,
 {\color{rossoCP3} New physics explanations of $a_\mu$ in light of the FNAL muon $g-2$ measurement},
[arXiv:2104.03691 [hep-ph]].


\bibitem{Aboubrahim:2021rwz}
A.~Aboubrahim, M.~Klasen and P.~Nath,
 {\color{rossoCP3} What Fermilab $(g-2)_{\mu}$ experiment tells us about discovering SUSY at HL-LHC and HE-LHC},
[arXiv:2104.03839 [hep-ph]].

\bibitem{Bhattacharya:2021ggm}
B.~Bhattacharya, A.~Datta, D.~Marfatia, S.~Nandi and J.~Waite,
 {\color{rossoCP3} Axion-like particles resolve the $B \to \pi K$ and $g-2$ anomalies},
[arXiv:2104.03947 [hep-ph]].



\bibitem{Kawamura:2021ygg}
J.~Kawamura and S.~Raby,
 {\color{rossoCP3} $\ge 4 \mu$ signal from a vector-like lepton decaying to a muon-philic $Z^\prime$ boson at the LHC},
[arXiv:2104.04461 [hep-ph]].

\bibitem{Baer:2021aax}
H.~Baer, V.~Barger and H.~Serce,
 {\color{rossoCP3} Anomalous muon magnetic moment, supersymmetry, naturalness, LHC search limits and the landscape},
[arXiv:2104.07597 [hep-ph]].


\bibitem{Buras:2021btx}
A.~J.~Buras, A.~Crivellin, F.~Kirk, C.~A.~Manzari and M.~Montull,
 {\color{rossoCP3} Global Analysis of Leptophilic Z' Bosons},
[arXiv:2104.07680 [hep-ph]].






  

\bibitem{Gerasimov:1965et}
  S.~B.~Gerasimov,
   {\color{rossoCP3} A sum rule for magnetic moments and the damping of the nucleon magnetic
  moment in nuclei},
  Sov.\ J.\ Nucl.\ Phys.\  {\bf 2}, 430 (1966)
  [Yad.\ Fiz.\  {\bf 2}, 598 (1966)].

\bibitem{Drell:1966jv}
  S.~D.~Drell and A.~C.~Hearn,
   {\color{rossoCP3} Exact sum rule for nucleon magnetic moments},
  Phys.\ Rev.\ Lett.\  {\bf 16}, 908 (1966).


\bibitem{Jaffe:1987hv}
  R.~L.~Jaffe and Z.~Ryzak,
   {\color{rossoCP3} Constraints from the Drell-Hearn-Gerasimov sum rule in chiral models of 
  composite fermions},
  Phys.\ Rev.\  D {\bf 37}, 2015 (1988).


\bibitem{Brodsky:1980zm}
  S.~J.~Brodsky and S.~D.~Drell,
   {\color{rossoCP3} The anomalous magnetic moment and limits on fermion substructure},
  Phys.\ Rev.\  D {\bf 22}, 2236 (1980).

\bibitem{Goldberg:1981gc}
  H.~Goldberg,
  {\color{rossoCP3} Bounds on $e^+ e^- \to l^* \bar l$ and $lp \to l^* X$ ($l^*$ = excited lepton) and prospects for visible $l^*$ tracks in cosmic ray emulsion events},
  Phys.\ Rev.\  D {\bf 24}, 1991 (1981).

\bibitem{Goldberg:1999gc} 
  H.~Goldberg,
   {\color{rossoCP3} Gravitons and the Drell-Hearn-Gerasimov sum rule: Support for large extra dimensions?},
  Phys.\ Lett.\ B {\bf 472}, 280 (2000)
  [hep-ph/9904318].




\bibitem{Brodsky:1995fj} 
  S.~J.~Brodsky and I.~Schmidt,
   {\color{rossoCP3} Classical photoabsorption sum rules},
  Phys.\ Lett.\ B {\bf 351}, 344 (1995)
  [hep-ph/9502416].

\bibitem{Mangano:1990by}
M.~L.~Mangano and S.~J.~Parke,
  {\color{rossoCP3}  Multiparton amplitudes in gauge theories},
Phys. Rept. \textbf{200} (1991), 301-367
doi:10.1016/0370-1573(91)90091-Y
[arXiv:hep-th/0509223 [hep-th]].


\bibitem{Dixon:1996wi}
L.~J.~Dixon,
  {\color{rossoCP3}  Calculating scattering amplitudes efficiently},
[arXiv:hep-ph/9601359 [hep-ph]].
  
\bibitem{Taylor:2017sph}
T.~R.~Taylor,
 {\color{rossoCP3}   A Course in Amplitudes},
Phys. Rept. \textbf{691} (2017), 1-37
doi:10.1016/j.physrep.2017.05.002
[arXiv:1703.05670 [hep-th]].



\bibitem{Sirunyan:2019vgj}
A.~M.~Sirunyan \textit{et al.} [CMS Collaboration],
  {\color{rossoCP3}  Search for high mass dijet resonances with a new background prediction method in proton-proton collisions at $\sqrt{s} =$ 13 TeV},
JHEP \textbf{05} (2020), 033
doi:10.1007/JHEP05(2020)033
[arXiv:1911.03947 [hep-ex]].

\bibitem{Sirunyan:2021khd}
A.~M.~Sirunyan \textit{et al.} [CMS Collaboration],
 {\color{rossoCP3}  Search for resonant and nonresonant new phenomena in high-mass dilepton final states at $\sqrt{s} = $ 13 TeV},
[arXiv:2103.02708 [hep-ex]].

\bibitem{Accomando:2010fz}
E.~Accomando, A.~Belyaev, L.~Fedeli, S.~F.~King and C.~Shepherd-Themistocleous,
 {\color{rossoCP3}  $Z'$ physics with early LHC data},
Phys. Rev. D \textbf{83} (2011), 075012
doi:10.1103/PhysRevD.83.075012
[arXiv:1010.6058 [hep-ph]].


\bibitem{Altarelli:1989ff}
G.~Altarelli, B.~Mele and M.~Ruiz-Altaba,
 {\color{rossoCP3}  Searching for new heavy vector bosons in $p \bar{p}$ colliders},
Z. Phys. C \textbf{45} (1989), 109
[erratum: Z. Phys. C \textbf{47} (1990), 676]
doi:10.1007/BF01556677

\bibitem{Antoniadis:1993jp}
I.~Antoniadis and K.~Benakli,
{\color{rossoCP3}  Limits on extra dimensions in orbifold compactifications of superstrings},
Phys. Lett. B \textbf{326} (1994), 69-78
doi:10.1016/0370-2693(94)91194-0
[arXiv:hep-th/9310151 [hep-th]].


\bibitem{Ali:2021xlw}
H.~Al Ali, N.~Arkani-Hamed, I.~Banta, S.~Benevedes, D.~Buttazzo, T.~Cai, J.~Cheng, T.~Cohen, N.~Craig and M.~Ekhterachian, \textit{et al.}
 {\color{rossoCP3}  The Muon Smasher's Guide},
[arXiv:2103.14043 [hep-ph]].





\end{thebibliography}
\end{document}